\documentclass[oneside,11pt]{article}
\usepackage{amssymb}
\usepackage{amsfonts}
\usepackage{graphicx}
\usepackage{amsmath}

\begin{document}

\title{Peridynamic Model for Single-Layer Graphene Obtained from Coarse Grained Bond Forces}
\author{Stewart A. Silling\footnote{Sandia National Laboratories, sasilli@sandia.gov}, Marta D'Elia\footnote{Sandia National Laboratories}, Yue Yu\footnote{Lehigh University},\\ Huaiqian You \footnote{Lehigh University}, M\"uge Fermen-Coker \footnote{Army Research Laboratory}}

\maketitle

\abstract{
An ordinary state-based peridynamic material model is proposed for single sheet graphene.
The model is calibrated using coarse grained molecular dynamics simulations.
The coarse graining method allows the dependence of bond force on bond length to
be determined, including the horizon.
The peridynamic model allows the horizon to be rescaled, providing a multiscale capability and
allowing for substantial reductions in computational cost compared with molecular dynamics. 
The calibrated peridynamic model is compared to experimental data on the deflection and perforation 
of a graphene monolayer
by an atomic force microscope probe.
}

\newcommand{\norm}[1]{\left\Vert#1\right\Vert}
\newcommand{\abs}[1]{\left\vert#1\right\vert}

\newcommand{\urho}{{\underline\rho}}
\newcommand{\bnu}{\boldsymbol{\nu}}
\newcommand{\betta}{ {\boldsymbol{\eta}} }
\newcommand{\bxi}{ {\boldsymbol{\xi}} }
\newcommand{\bzeta}{ {\boldsymbol{\zeta}} }
\newcommand{\bsigma}{\boldsymbol{\sigma}}
\newcommand{\btau}{ {\mbox{\boldmath$\tau$}} }
\newcommand{\onR}{{\quad{\rm on\Space}\R}}
\newcommand{\horizon}{\delta}
\newcommand{\dirac}{\Delta}
\newcommand{\rthree}{{\mathbb{R}^3}}
\newcommand{\rone}{{\mathbb{R}}}
\newcommand{\bexch}{ {\underline{\mbox{\boldmath$\delta$}} }}
\newcommand{\exch}{ {\underline{\delta}} }

\newcommand{\ubeta}{{\underline \beta}}
\newcommand{\utau}{{\underline \tau}}
\newcommand{\umu}{{\underline \mu}}
\newcommand{\uvarphi}{{\underline \varphi}}
\newcommand{\Oplus}{{{\mathcal{O}}^+}}
\newcommand{\cA}{{\mathcal{A}}}
\newcommand{\cB}{{\mathcal{B}}}
\newcommand{\cBP}{{\mathcal{B}\setminus \mathcal{P}}}
\newcommand{\cBp}{{\cB_\bp}}
\newcommand{\cBq}{{\cB_\bq}}
\newcommand{\calB}{{\mathcal{B}}}
\newcommand{\calC}{{\mathcal{C}}}
\newcommand{\calL}{{\mathcal{L}}}
\newcommand{\cP}{{\mathcal{P}}}
\newcommand{\R}{{\mathcal{R}}}
\newcommand{\Real}{{\mathbb{R}}}
\newcommand{\calV}{{\mathcal{V}}}
\newcommand{\cD}{{\mathcal{D}}}
\newcommand{\cE}{{\mathcal{E}}}
\newcommand{\cF}{{\mathcal{F}}}
\newcommand{\cH}{{\mathcal{H}}}
\newcommand{\cHx}{{\mathcal{H}_{\bf x}}}
\newcommand{\cHxp}{{\mathcal{H}_{\bf x'}}}
\newcommand{\cHq}{{\mathcal{H}_{\bf q}}}
\newcommand{\cO}{{\mathcal{O}^+}}
\newcommand{\ucO}{{\mathcal{\underline O}^+}}
\newcommand{\cQ}{{\mathcal{Q}}}
\newcommand{\calZ}{{\mathcal{Z}}}
\newcommand{\calCNpw}{{\calC^N_{\rm pw} }}
\newcommand{\Sr}{{{\mathcal{S}}_r}}
\newcommand{\bzero}{{\bf {0}}}
\newcommand{\uzero}{{\underline 0}}
\newcommand{\ubzero}{{\underline {\bf 0} }}
\newcommand{\ba}{{\bf a}}
\newcommand{\uba}{{\underline{\bf a}}}
\newcommand{\bA}{{\bf A}}
\newcommand{\uA}{{\underline A}}
\newcommand{\ua}{{\underline a}}
\newcommand{\uad}{{{\underline a}^{\rm d}}}
\newcommand{\ubA}{{\underline{\bf A}}}
\newcommand{\uubA}{{\underline{\mathbb{A}}}}
\newcommand{\bb}{{\bf b}}
\newcommand{\bB}{{\bf B}}
\newcommand{\uB}{{\underline B}}
\newcommand{\ub}{{\underline b}}
\newcommand{\ubb}{{\underline{\bf b}}}
\newcommand{\ubB}{{\underline{\bf B}}}
\newcommand{\uubB}{{\underline{\mathbb{B}}}}
\newcommand{\bc}{{\bf c}}
\newcommand{\bC}{{\bf C}}
\newcommand{\uC}{{\underline C}}
\newcommand{\uuC}{{\underline{C}}}
\newcommand{\uc}{{\underline c}}
\newcommand{\ubC}{{\underline{\bf C}}}
\newcommand{\uubC}{{\underline{\mathbb{C}}}}
\newcommand{\bD}{{\bf D}}
\newcommand{\uD}{{\underline D}}
\newcommand{\uuD}{{\underline{D}}}
\newcommand{\ud}{{\underline d}}
\newcommand{\uubD}{{\underline{\mathbb{D}}}}
\newcommand{\ucalE}{{\underline{\mathcal{E}}}}
\newcommand{\bh}{{\bf h}}
\newcommand{\bhp}{\mathbf{h}^\prime}
\newcommand{\bH}{{\bf H}}
\newcommand{\bI}{{\bf I}}
\newcommand{\uI}{{\underline 1}}
\newcommand{\ubI}{{\underline{\bf I}}}
\newcommand{\br}{{\bf r}}
\newcommand{\uubR}{{\underline{\mathbb{R}}}}
\newcommand{\ubV}{{\underline{\bf V}}}
\newcommand{\dt}{{\Delta t}}
\newcommand{\dV}{{\Delta V}}
\newcommand{\uv}{{\underline v}}
\newcommand{\dx}{{\Delta x}}
\newcommand{\dy}{{\Delta y}}
\newcommand{\dz}{{\Delta z}}
\newcommand{\bE}{{\bf E}}
\newcommand{\be}{{\bf e}}
\newcommand{\uubE}{{\underline{\mathbb{E}}}}
\newcommand{\ue}{{\underline e}}
\newcommand{\ued}{{{\underline e}^{\rm d}}}
\newcommand{\uede}{{{\underline e}^{\rm de}}}
\newcommand{\uedp}{{{\underline e}^{\rm dp}}}
\newcommand{\uei}{{{\underline e}^{\rm i}}}
\newcommand{\dotue}{{{\underline{\dot e}}}}
\newcommand{\dotuei}{{{\dotue}^{\rm i}}}
\newcommand{\dotued}{{{\dotue}^{\rm d}}}
\newcommand{\dotuedp}{{{\dotue}^{\rm dp}}}
\newcommand{\dotuede}{{{\dotue}^{\rm de}}}
\newcommand{\ucE}{{\underline{\cal E}}}
\newcommand{\uE}{{{\underline{E}}}}
\newcommand{\ubE}{{{\underline{\bf E}}}}
\newcommand{\bex}{{\bf e}_x}
\newcommand{\bey}{{\bf e}_y}
\newcommand{\bez}{{\bf e}_z}
\newcommand{\bff}{{\bf f}}
\newcommand{\bffp}{\mathbf{f}^\prime}
\newcommand{\bF}{{\bf F}}
\newcommand{\bFT}{{ {\bf F}^T }}
\newcommand{\ubF}{{\underline{\bf F}}}
\newcommand{\bg}{{\bf g}}
\newcommand{\bG}{{\bf G}}
\newcommand{\bgp}{\mathbf{g}^\prime}
\newcommand{\ug}{{\underline{g}}}
\newcommand{\uG}{{\underline{G}}}
\newcommand{\ubG}{{\underline{\bf G}}}
\newcommand{\uubG}{{\underline{\mathbb{G}}}}
\newcommand{\ubH}{{\underline{\bf H}}}
\newcommand{\cI}{{\cal I}}
\newcommand{\cJ}{{\cal J}}
\newcommand{\bJ}{{\bf J}}
\newcommand{\bj}{{\bf j}}
\newcommand{\bK}{{\bf K}}
\newcommand{\bk}{{\bf k}}
\newcommand{\uk}{{\underline k}}
\newcommand{\uK}{{\underline K}}
\newcommand{\uubK}{{\underline{\mathbb{K}}}}
\newcommand{\bL}{{\bf L}}
\newcommand{\bLu}{{\bf L}_\bu}
\newcommand{\bLv}{{\bf L}_\bv}
\newcommand{\bmm}{{\bf m}}
\newcommand{\bM}{{\bf M}}
\newcommand{\uM}{{\underline M}}
\newcommand{\ubM}{{\underline{\bf M}}}
\newcommand{\bn}{{\bf n}}
\newcommand{\cN}{{\mathcal N}}
\newcommand{\bN}{{\bf N}}
\newcommand{\uN}{{\underline N}}
\newcommand{\ubN}{{\underline{\bf N}}}
\newcommand{\bone}{{\bf 1}}
\newcommand{\ubone}{{\underline{\bf 1}}}
\newcommand{\uone}{{\underline{1}}}
\newcommand{\bp}{{\bf p}}
\newcommand{\bP}{{\bf P}}
\newcommand{\sfP}{{\mathsf{P}}}
\newcommand{\uP}{{\underline P}}
\newcommand{\up}{{\underline p}}
\newcommand{\ubP}{{\underline{\bf P}}}
\newcommand{\uubP}{{\underline{\mathbb{P}}}}
\newcommand{\bq}{{\bf q}}
\newcommand{\bQ}{{\bf Q}}
\newcommand{\uQ}{{\underline Q}}
\newcommand{\ubQ}{{\underline{\bf Q}}}
\newcommand{\ubr}{{\underline{\bf r}}}
\newcommand{\bR}{{\bf R}}
\newcommand{\sfR}{{\mathsf{R}}}
\newcommand{\ur}{{\underline r}}
\newcommand{\sfS}{{\mathsf{S}}}
\newcommand{\bS}{{\bf S}}
\newcommand{\bs}{{\bf s}}
\newcommand{\uS}{{\underline S}}
\newcommand{\ubs}{{\underline{\bf s}}}
\newcommand{\ubS}{{\underline{\bf S}}}
\newcommand{\uubS}{{\underline{\mathbb{S}}}}
\newcommand{\cs}{{\cal s}}
\newcommand{\cS}{{\cal S}}
\newcommand{\cSd}{{\cal S^{\rm d}}}
\newcommand{\bt}{{\bf t}}
\newcommand{\btp}{\mathbf{t}^\prime}
\newcommand{\bT}{{\bf T}}
\newcommand{\uT}{{\underline T}}
\newcommand{\ut}{{\underline t}}
\newcommand{\dotut}{{{\underline{\dot t}}}}
\newcommand{\dotuti}{{{\dotut}^{\rm i}}}
\newcommand{\dotutd}{{{\dotut}^{\rm d}}}
\newcommand{\ubT}{{\underline{\bf T}}}
\newcommand{\utd}{{{\underline t}^{\rm d}}}
\newcommand{\uti}{{{\underline t}^{\rm i}}}
\newcommand{\bu}{{\bf u}}
\newcommand{\tu}{{\tilde u}}
\newcommand{\ubu}{{\underline{\bf u}}}
\newcommand{\uu}{{\underline{u}}}
\newcommand{\bU}{{\bf U}}
\newcommand{\uU}{{\underline U}}
\newcommand{\ubU}{{\underline{\bf U}}}
\newcommand{\bv}{{\bf v}}
\newcommand{\bV}{{\bf V}}
\newcommand{\bvp}{\mathbf{v}^\prime}
\newcommand{\cL}{{\cal L}}
\newcommand{\cR}{{\mathcal R}}
\newcommand{\cV}{{\cal V}}
\newcommand{\bw}{{\bf w}}
\newcommand{\bwp}{\mathbf{w}^\prime}
\newcommand{\bW}{{\bf W}}
\newcommand{\brw}{{\breve w}}
\newcommand{\uW}{{\underline W}}
\newcommand{\ubW}{{\underline{\bf W}}}
\newcommand{\bx}{{\bf x}}
\newcommand{\bxp}{\mathbf{x}^\prime}
\newcommand{\bX}{{\bf X}}
\newcommand{\uX}{{\underline X}}
\newcommand{\ux}{{\underline x}}
\newcommand{\ubX}{{\underline{\bf X}}}
\newcommand{\by}{{\bf y}}
\newcommand{\byp}{\mathbf{y}^\prime}
\newcommand{\bY}{{\bf Y}}
\newcommand{\uY}{{\underline Y}}
\newcommand{\uy}{{\underline y}}
\newcommand{\ubY}{{\underline{\bf Y}}}
\newcommand{\cYd}{{ {\textsf Y} }}
\newcommand{\bz}{{\bf z}}
\newcommand{\bZ}{{\bf Z}}
\newcommand{\uZ}{{\underline Z}}
\newcommand{\uz}{{\underline z}}
\newcommand{\ubZ}{{\underline{\bf Z}}}
\newcommand{\ubz}{{\underline{\bf z}}}
\newcommand{\motion}{{\by}}
\newcommand{\bchi}{{\boldsymbol \chi}}
\newcommand{\bkappa}{{\boldsymbol \kappa}}
\newcommand{\blambda}{{\boldsymbol \lambda}}
\newcommand{\uomega}{{\underline \omega}}
\newcommand{\uomegas}{{{\underline \omega}_s}}
\newcommand{\uphi}{{\underline \phi}}
\newcommand{\ubphi}{{\underline{\bf \phi}}}
\newcommand{\uPhi}{{\underline \Phi}}
\newcommand{\ubPhi}{{\underline{\bf \Phi}}}
\newcommand{\Phiu}{{\Phi_\bu}}
\newcommand{\bPhi}{{\boldsymbol{\Phi}}}
\newcommand{\dotPhiu}{{{\dot\Phi}_\bu}}
\newcommand{\bpsi}{{\boldsymbol{\psi}}}
\newcommand{\bpsidef}{ {\boldsymbol{\psi}}_{\text{def}} }
\newcommand{\upsi}{{\underline \psi}}
\newcommand{\uPsi}{{\underline \Psi}}
\newcommand{\ubPsi}{{\underline{\bf \Psi}}}
\newcommand{\utheta}{{\underline \theta}}
\newcommand{\Wu}{{W_\bu}}
\newcommand{\eps}{{\epsilon}}
\newcommand{\epsinv}{{\eps^{-1}}}
\newcommand{\beps}{{\boldsymbol{\epsilon}}}
\newcommand{\vareps}{{\varepsilon}}
\newcommand{\epsoneone}{{\epsilon_{11}}}
\newcommand{\epstwotwo}{{\epsilon_{22}}}
\newcommand{\epsthreethree}{{\epsilon_{33}}}
\newcommand{\Space}{{\ }}
\newcommand{\prl}{{\parallel}}
\newcommand{\basis}{{  \{\be_1,\be_2,\be_3\}  }}
\newcommand{\onehalf}{{\frac12}}
\newcommand{\buone}{{{\bf u}^{(1)}}}
\newcommand{\butwo}{{{\bf u}^{(2)}}}
\newcommand{\rplus}{{\R^+}}
\newcommand{\rminus}{{\R^-}}
\newcommand{\Arccos}{{\cos^{-1}}}
\newcommand{\Arcsin}{{\sin^{-1}}}
\newcommand{\uarrows}{{U}}
\newcommand{\Uhat}{{\hat{U}}}
\newcommand{\aovery}{{ \left({a\over y}\right) }}
\newcommand{\Ex}{{ \;{\rm Ex}\; }}
\newcommand{\Dir}{{ {\rm Dir}\; }}
\newcommand{\nablad}{{\nabla^{\rm d}}}
\newcommand{\la}{{\langle}}
\newcommand{\ra}{{\rangle}}
\newcommand{\frechy}{{\nabla_{\ubY}}}
\newcommand{\frechu}{{\nabla_{\ubu}}}
\newcommand{\frechU}{{\nabla_{\ubU}}}
\newcommand{\frecht}{{\nabla_{\ubT}}}
\newcommand{\bKinv}{\bK^{-1}}
\newcommand{\Kinv}{K^{-1}}
\newcommand{\qor}{\quad{\textrm{or}}\quad}
\newcommand{\dv}{{\;dV}}
\newcommand{\dvp}{{\;dV^\prime}}
\newcommand{\dvx}{{\;{\textrm{d}}V_\bx}}
\newcommand{\dvbp}{{\;dV_\bp}}
\newcommand{\dvbq}{{\;{\textrm{d}}V_\bq}}
\newcommand{\dvxp}{{\;dV_{\bx^\prime}}}
\newcommand{\dvxi}{{\;{\textrm{d}}V_\bxi}}
\newcommand{\dvzeta}{{\;{\textrm{d}}V_\bzeta}}
\newcommand{\dveta}{{\;{\textrm{d}}V_\betta}}
\newcommand{\dvz}{{\;dV_\bz}}
\newcommand{\frechetY}{{ \mathop{\nabla}_{\small\ubY} }}
\newcommand{\frechetdY}{{ \mathop{\nabla}_{\small\dot\ubY} }}
\newcommand{\acos}{{ \cos^{-1} }}
\newcommand{\adj}{^\dagger}
\newcommand{\calE}{{ \mathcal{E} }}
\newcommand{\calK}{{ \mathcal{K} }}
\newcommand{\calQ}{{ \mathcal{Q} }}
\newcommand{\calT}{{ \mathcal{T} }}
\newcommand{\calU}{{ \mathcal{U} }}
\newcommand{\calW}{{ \mathcal{W} }}
\newcommand{\calY}{{ \mathcal{Y} }}
\newcommand{\Wabs}{{ \calW_{\mathrm{abs}} }}
\newcommand{\Wsup}{{ \calW_{\mathrm{sup}} }}
\newcommand{\pabs}{{ p_{\mathrm{abs}} }}
\newcommand{\Shard}{{ {\cS}_{\mathrm{hard}} }}
\newcommand{\Ssoft}{{ {\cS}_{\mathrm{soft}} }}
\newcommand{\Jinv}{{ J^{-1} }}
\newcommand{\rhodef}{{ {\rho}_{\text{def}} }}
\newcommand{\dotrhodef}{{ \dot{\rho}_{\text{def}} }}
\newcommand{\hatrhodef}{{ \hat{\rho}_{\text{def}} }}
\newcommand{\dothatrhodef}{{ \dot{\hat{\rho}}_{\text{def}} }}
\newcommand{\rholocal}{{ {\rho}_{\text{local}} }}
\newcommand{\dotrholocal}{{ \dot{\rho}_{\text{local}} }}
\newcommand{\bvdef}{{ {\bv}_{\text{def}} }}
\newcommand{\imag}{{ {\text{Im}} }}
\newcommand{\real}{{ {\text{Re}} }}
\newcommand{\sgn}{{ {\text{sgn}} }}
\newcommand{\erf}{{ {\text{erf}} }}
\newcommand{\Tr}{{ {\text{Tr}}\, }}
\newcommand{\Iso}{{ {\text{Iso}}\, }}
\newcommand{\Dev}{{ {\text{Dev}}\, }}
\newcommand{\dd}{{ {\text{d}} }}
\newcommand{\dns}{{ {\text{DNS}} }}
\newcommand{\UX}{{ {\text{UX}} }}
\newcommand{\IE}{{ {\text{IE}} }}

\section{Introduction}\label{sec-intro}
Molecular dynamics has made enormous advances in capabilities through better algorithms,
better interatomic potentials, and improvements in computational power.
However, the use of molecular dynamics directly to treat the deformation and failure of
materials at the mesoscale is still largely beyond reach.
At the mesoscale and above, a continuum model of mechanics is still required in practice.
The question then arises of how molecular dynamics can be used in deriving and calibrating
appropriate continuum models.
This paper addresses the question of how to use molecular dynamics to obtain a peridynamic
material model that is able to treat material nonlinearity and the nucleation and growth of fractures.

To accomplish this, a coarse graining method is described below that maps interatomic forces
into larger-scale degrees of freedom.
The coarse graining method starts with a definition of these degrees of freedom as the
mean atomic displacements weighted by a smoothing function.
It is shown that the coarse grained displacements obey a nonlocal evolution law, which
is the peridynamic equation of motion.

The coarse graining process provides peridynamic bond forces among the coarse grained nodes
that are then used to calibrate a material model.
The bond forces can include long-range interactions, if these are present in the atomic system.
They also reflect any initial distribution of defects.

In the present application of single-sheet graphene, a nonlinear ordinary state-based material
model is found to adequately represent the deformation and failure of the material.
As a molecular dynamics (MD) model of graphene is stretched, the interatomic forces become weaker, and the material
fails.
This process of failure is accelerated by higher temperatures in the MD model, which also affect
the elastic response.
All of these features are reflected in the coarse grained bond forces, so they are carried over
to the peridynamic continuum model after calibration.

The calibrated peridynamic model reproduces the nucleation of damage due to deformation in a
specimen that is initially undamaged.
In principle, the model can be applied within the process zone of a growing crack.
However, with the objective of scaling up the material model to much larger length scales,
it is necessary to include a separate bond breakage criterion that reflects the energy balance in
brittle crack growth without the need to model the process zone in detail.
To treat this,
the peridynamic material model is augmented by a separate bond breakage criterion that approximates
the Griffith criterion for growing cracks in a brittle material.

The literature on graphene is voluminous, and only the papers that are the most relevant to
the present work are summarized here.
Much of what is known about the mechanical properties of graphene is based on MD simulations.
Jiang, Wang, and Li used MD to predict the Young's modulus in graphene, including the effects of
temperature and sample size \cite{jiang09}.
A number of MD studies have treated the
effect of defects on the mechanical and thermal properties of graphene
\cite{ni10,jing12,ansari12,mortazavi13,he14}.
Sakhee-Pour \cite{sakhaee09} and
Javvaji et al. investigated the effects of lattice orientation and sample size on the strength
of graphene \cite{javvaji16}.
Most of these papers, as well as the present paper, treat only the two-dimensional response
of graphene.
However, 3D MD simulations have also been applied to the wrinkling and crumpling of graphene
sheets, for example \cite{becton15}.
MD has also been used to study the mechanical properties of  polycrystalline graphene, for
example \cite{yi13,chen15}.
A comprehensive review of the literature on the fracture of graphene, much of which uses MD,
can be found in \cite{zhang15}.
A review of the literature on experimental and theoretical graphene mechanics is available in \cite{cao18}.

Continuum modeling of single-layer graphene has included the use of finite elements with an elastic 
material model, for example \cite{hemmasizadeh08,scarpa10}.
A summary of the literature on the equivalent linear elastic properties of graphene sheets is
given by Reddy et al. \cite{reddy06} and by Shi et al. \cite{shi14}.
Finite element analysis including aspects of fracture mechanics has been applied to graphene sheets
\cite{tsai10}.
A hyperelastic continuum material model that includes nonlinearity at large strains was developed
by Xu et al. using density functional theory \cite{xu12}.
An up-to-date review of the literature on finite element modeling of graphene is given
by Chandra et al. \cite{chandra20}.
Nonlocality has been studied in connection to the buckling of single-layer graphene 
\cite{pradhan09,pradhan10,asemi14}
and is potentially important in the modeling of multilayer graphene,
partly due to the long-range interaction forces between layers.

Liu et al. \cite{liu18} developed an ordinary state-based peridynamic model for single-layer 
graphene that is calibrated using strain energy densities obtained from MD.
Nonlinearity in the stress-strain response is incorporated by including a cubic dependence
of strain energy density on strain.
This method reproduces the stress-strain curves predicted by MD and, when a critical strain
bond breakage criterion is used, also captures the main features of dynamic fracture that are
seen in MD.
The method in \cite{liu18} does not address the dependence of bond force on bond length,
which is treated in the present work.

Other applications of peridynamics to graphene include the work of
Martowicz et al. \cite{martowicz15}, which uses a peridynamic model of graphene nanoribbons to
reproduce wave dispersion.
Diyaroglu et al. \cite{diyaroglu19} apply peridynamics to the wrinkling of graphene membranes,
including thermal expansion.
Liu et al. \cite{liu20} present a bond-based treatment of the
effects of lattice orientation on the strength of graphene sheets in different directions.
A bond-based material model has been applied to the perforation of multilayer graphene
by micrometer-scale projectiles \cite{silling21b}.

In Section~\ref{sec-homogpd} of the present paper, an upscaling method is presented that
provides coarse grained bond forces that are consistent with the momentum balance for the smoothed displacement
variable.
Section~\ref{sec-example} presents an example of coarse graining in a linear small-scale system that
involves long-range forces.
This section also describes the fitting of a peridynamic material model to the coarse grained forces.
Section~\ref{sec-graphene} extends the method to the nonlinear response of graphene, including
the nucleation of damage.
Section~\ref{sec-breakage} describes how a critical bond strain damage criterion can be
combined with the peridynamic model to reproduce the growth of cracks.
In Section~\ref{sec-scale} it is shown how changes in the horizon can be applied to the model with
appropriate scaling of the parameters.
Comparison of a simulation using the new material model for graphene with experimental data on the rupture of nanoscale
membranes is presented in Section~\ref{sec-exper}.
Concluding remarks and ideas for future work are given in Section~\ref{sec-disc}.

\section{Coarse graining of an atomic scale model}\label{sec-homogpd}
This section describes a method for obtaining a larger-scale discretized model from an MD model.
The discussion specializes a more general method described in \cite{silling21a} to the case of 
discrete nodes.
The general approach is to first define the coarse grained displacements in terms
of a weighted average of the microscale displacements.
This definition leads to a linear momentum balance for the coarse grained  displacements
that is a consequence of the momentum balance for the atoms.
The coarse grained momentum balance has the form of the discretized peridynamic equation of motion.
The bond forces in this peridynamic expression are derived from the atomic scale forces.
How to determine a material model for the coarse grained bond forces is considered
in Section~\ref{sec-graphene}.

Consider a molecular dynamics model of a crystal composed of $N_a$ atoms. 
Over time, each atom $\alpha$ interacts with the same set of its neighbors $\cH_\alpha$.
The mass and displacement of each atom are denoted by $M_\alpha$ and $\bU_\alpha(t)$ respectively.
The atoms interact through some given interatomic potential.
The resulting force that
atom $\beta$ exerts on $\alpha$ is denoted by $\bF_{\beta\alpha}(t)$.
These interatomic forces obey the following antisymmetry relation:
\begin{equation}
  \bF_{\alpha\beta}(t)=-\bF_{\beta\alpha}(t)
\label{eqn-Fsym}
\end{equation}
for all $t$.
The forces are not necessarily parallel to the relative position vector between $\alpha$ and $\beta$.
Each atom is also subjected to a prescribed external force $\bB_\alpha(t)$.
The atoms obey Newton's second law:
\begin{equation}
  M_\alpha \ddot\bU_\alpha(t)=\sum_{\beta\in\cH_\alpha}\bF_{\beta\alpha}(t) +\bB_\alpha(t).
\label{eqn-newton}
\end{equation}
To coarse grain the molecular dynamics model, let $\bx_i$, $i=1,2,\dots,N_c$ denote the reference positions  
of the coarse grained degrees of freedom.
Let $\bu_i(t)$ denote the displacements at each such position, to be defined below.
For each $\bx_i$, define smoothing weights $\omega_i^\alpha$.
These weights are normalized such that for any atom $\alpha$,
\begin{equation}
  \sum_{i=1}^{N_c} \omega_i^\alpha=1.
\label{eqn-omeganorm}
\end{equation}
Equation \eqref{eqn-omeganorm} implies that each atom is covered by at least one smoothing function.
All of the weights are limited to a support of radius $R$:
\begin{equation}
  |\bx_i-\bX_\alpha|>R \quad\implies\quad \omega_i^\alpha=0
\label{eqn-Rdef}
\end{equation}
for any $i$ and $\alpha$, where $R$ is independent of $i$ and $\alpha$.
Define the coarse grained masses and external loads by
\begin{equation}
  m_i= \sum_{\alpha=1}^{N_a} \omega_i^\alpha M_\alpha, \qquad
  \bb_i(t)= \sum_{\alpha=1}^{N_a} \omega_i^\alpha \bB_\alpha(t).
\label{eqn-mbdef}
\end{equation}
It is assumed for convenience that $m_i>0$ for all $i$, that is, for every $i$, there is some atom
$\alpha$ such that $\omega_i^\alpha>0$.
Define the coarse grained displacements by
\begin{equation}
  \bu_i(t)= \frac{1}{m_i}\sum_{\alpha=1}^{N_a} \omega_i^\alpha M_\alpha \bU_\alpha(t).
\label{eqn-udef}
\end{equation}
Thus, the coarse grained displacements are weighted by mass as well as $\omega_i^\alpha$.

Next, the evolution equation for the coarse grained displacements will be derived.
Taking the second time derivative of \eqref{eqn-udef} yields
\begin{equation}
  m_i\ddot\bu_i(t)= \sum_{\alpha=1}^{N_a} \omega_i^\alpha M_\alpha \ddot\bU_\alpha(t).
\label{eqn-usec}
\end{equation}
From \eqref{eqn-newton} and \eqref{eqn-usec},
\begin{equation}
  m_i\ddot\bu_i(t)= \sum_{\alpha=1}^{N_a} \omega_i^\alpha \left[\sum_{\beta=1}^{N_a}\bF_{\beta\alpha}(t) +\bB_\alpha(t)\right].
\label{eqn-useci}
\end{equation}
For any atom $\beta$, the normalization requirement \eqref{eqn-omeganorm} implies that
\begin{equation}
  \sum_{j=1}^{N_c} \omega_j^\beta=1.
\label{eqn-betanorm}
\end{equation}
Combining \eqref{eqn-useci} and \eqref{eqn-betanorm}, and using the second equation in \eqref{eqn-mbdef},
\begin{equation}
  m_i\ddot\bu_i(t)= \sum_{\alpha=1}^{N_a} \omega_i^\alpha \left[\sum_{\beta=1}^{N_a}\bF_{\beta\alpha}(t)\sum_{j=1}^{N_c} \omega_j^\beta\right] +\bb_i(t).
\label{eqn-usecii}
\end{equation}
Rearranging \eqref{eqn-usecii} leads to 
\begin{equation}
  m_i\ddot\bu_i(t)= \sum_{j=1}^{N_c} \bff_{ji}(t) +\bb_i(t)
\label{eqn-usecpd}
\end{equation}
where the {\emph{pairwise bond force}} is defined by
\begin{equation}
  \bff_{ji}(t)= \sum_{\alpha=1}^{N_a} \sum_{\beta=1}^{N_a} \omega_i^\alpha \omega_j^\beta \bF_{\beta\alpha}(t).
\label{eqn-ffdef}
\end{equation}
Using \eqref{eqn-Fsym} and interchanging the summation variables $\alpha$ and $\beta$, it follows immediately from \eqref{eqn-ffdef} that
\begin{equation}
  \bff_{ij}(t)=-\bff_{ji}(t)
\label{eqn-fsym}
\end{equation}
for all $i$, $j$, and $t$.

\begin{figure} 
\centering
\includegraphics[width=1.2\textwidth]{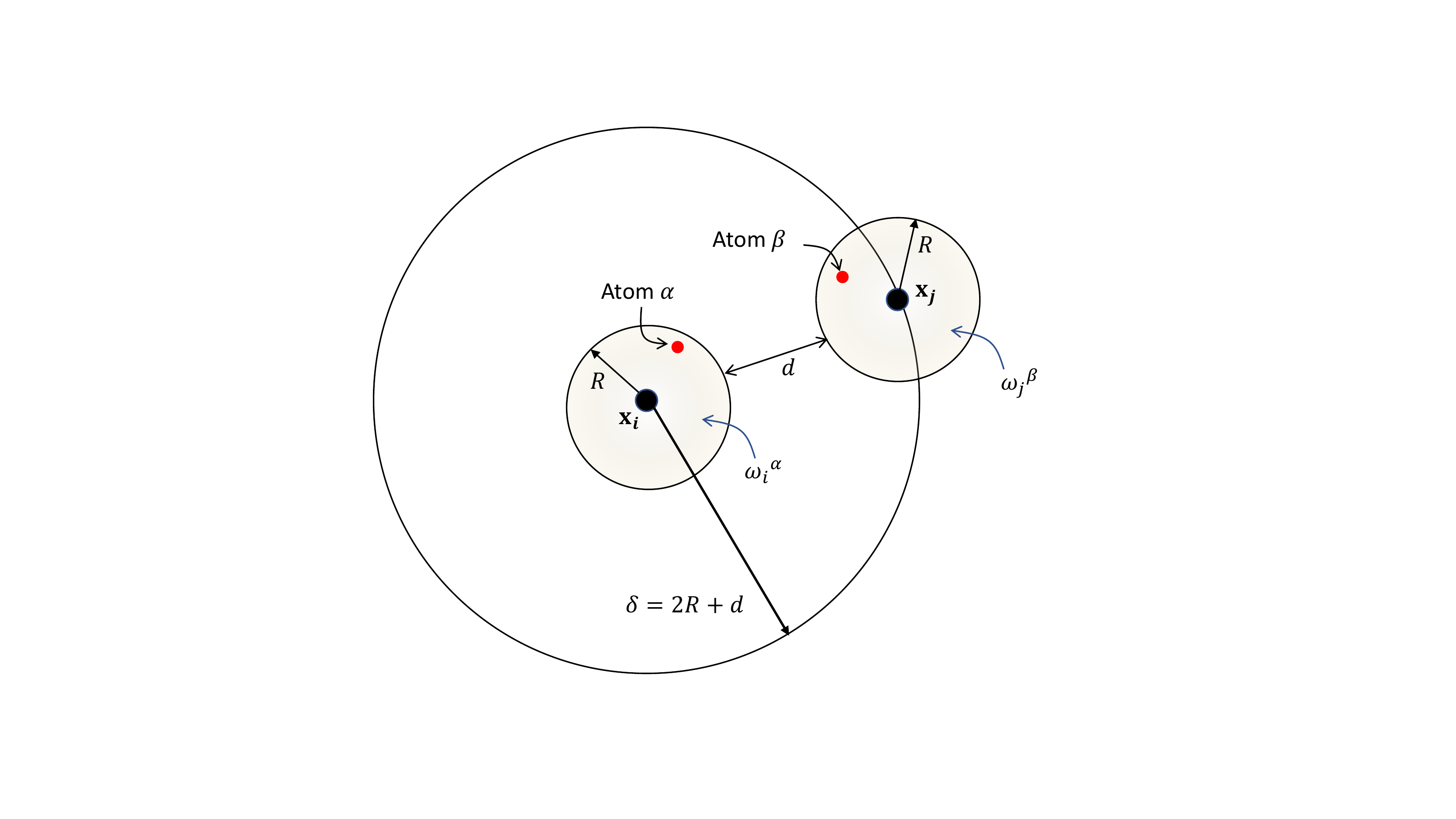}
\caption{The horizon is determined by the weight function radius and the cutoff distance for 
the interatomic potential.}
\label{fig-horizon}
\end{figure}

Suppose that the underlying interatomic potential has a cutoff distance $d$:
\begin{equation}
  |\bX_\beta-\bX_\alpha|>d \quad\implies\quad \bF_{\beta\alpha}(t)=\bzero
\label{eqn-ddef}
\end{equation}
for all $\alpha$, $\beta$, and $t$.
As suggested by Figure~\ref{fig-horizon},
\eqref{eqn-Rdef}, \eqref{eqn-ffdef}, and \eqref{eqn-ddef} imply that 
\begin{equation}
  |\bx_j-\bx_i|>\delta \quad\implies\quad \bff_{ji}(t)=\bzero
\label{eqn-deltadefi}
\end{equation}
for all $i$, $j$, and $t$, where $\delta$ is the {\emph{horizon}} defined by
\begin{equation}
  \delta=2R+d.
\label{eqn-deltadef}
\end{equation}
So, $\delta$ is the cutoff distance for coarse grained bond force interactions.

The definition of $\bff_{ji}$ given by \eqref{eqn-ffdef} does not, in itself, provide a viable material model
for the coarse grained model.
Such a material model would relate the pairwise bond forces to the coarse grained displacements, not
to the interatomic forces, which would be unknown in a coarse grained model.
However, \eqref{eqn-ffdef} does provide a means to calibrate a prescribed material model, as will be
demonstrated in the next section.

\section{Example}\label{sec-example}
Consider a square lattice of particles in 2D, with spacing $\ell=1$ and layer thickness $\tau=1$.
The mass of each particle is $M_\alpha=1$.
The particles interact according to the following hypothetical model:
\begin{equation}
  \bF_{\beta\alpha} = \left\lbrace \begin{array}{ll}
      (B\bM_{\beta\alpha}) e^{-|\bX_\beta-\bX_\alpha|/d}  \quad & {\mathrm{if}}\;  |\bX_\beta-\bX_\alpha|\le d, \\
      0  & {\mathrm{otherwise}} \\
    \end{array}\right.
\label{eqn-sqlatt}
\end{equation}
where
\begin{equation}
      \bM_{\beta\alpha} =  \frac{ \bX_\beta-\bX_\alpha} {|\bX_\beta-\bX_\alpha|} 
\label{eqn-Mba}
\end{equation}
and $B=0.90915$, $d=10$.
Thus, long-range interactions are present up to 10 interatomic distances.

\begin{figure} 
\centering
\includegraphics[width=1.2\textwidth]{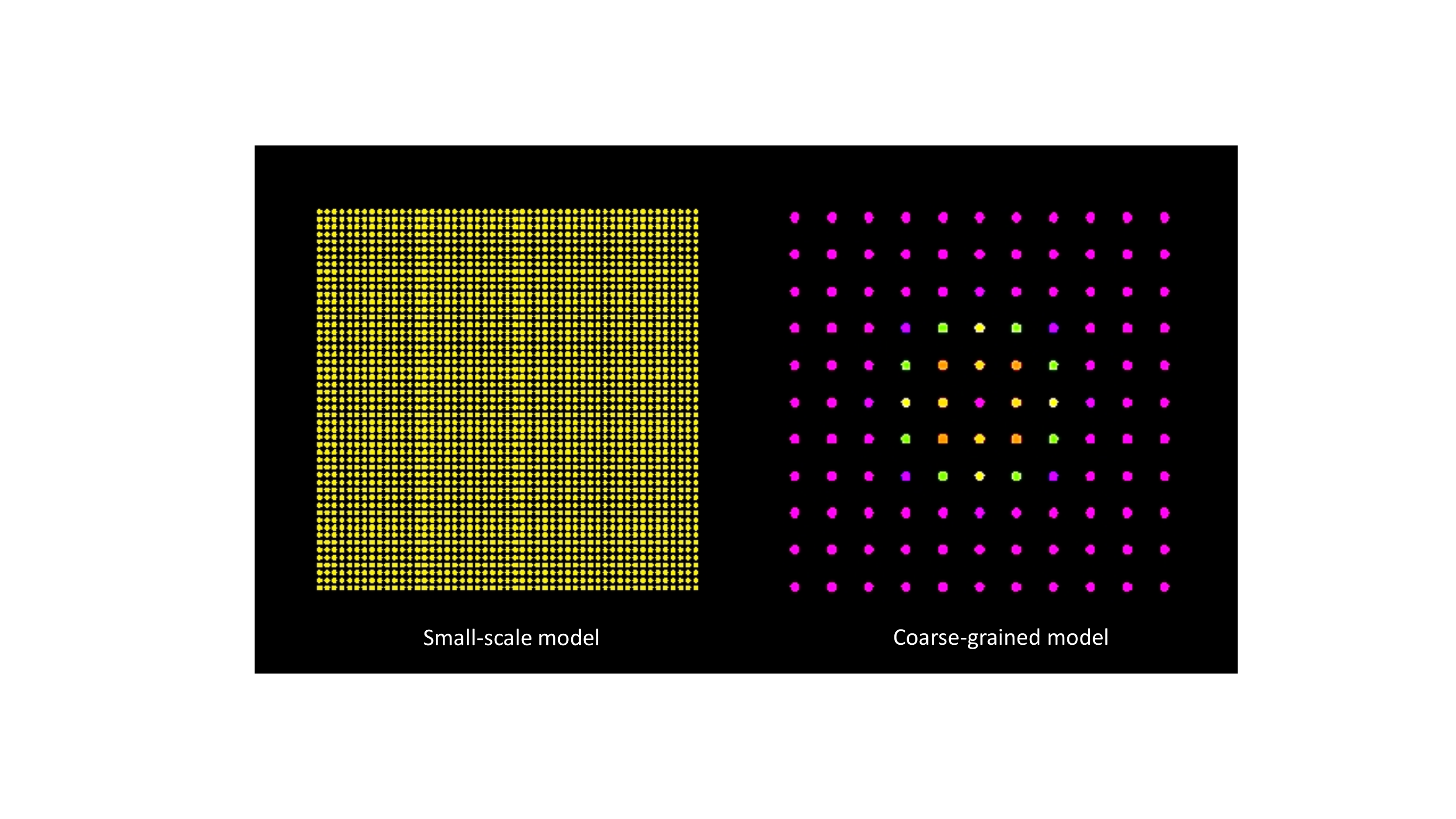}
\caption{Coarse graining example.
Left: original small-scale grid.
Right: coarse grained nodes. Colors show the force in bonds connected to the center CG node.
}
\label{fig-lrsquare}
\end{figure}

The coarse grained nodes are on a square lattice with a spacing of $h=5$ (Figure~\ref{fig-lrsquare}).
The smoothing functions are defined with the help of the cone-shaped function $S$ given by
\begin{equation}
  S(\bz)=\max\left\{0, 1-\frac{1}{R}\sqrt{z_1^2+z_2^2}  \right\}
\label{eqn-Sdef}
\end{equation}
where $z_1$ and $z_2$ are the components of the vector $\bz$ in the plane
and where $R$ is the radius of the cone.
In this example, $R=2h$.
The weighting functions $\omega_i^\alpha$ are given by
\begin{equation}
  \omega_i^\alpha=\frac{ S(\bx_i-\bX_\alpha)} {\sum_j S(\bx_j-\bX_\alpha)}
\label{eqn-Ommd}
\end{equation}
which is designed to satisfy the normalization \eqref{eqn-omeganorm}.

\begin{figure} 
\centering
\includegraphics[width=1.2\textwidth]{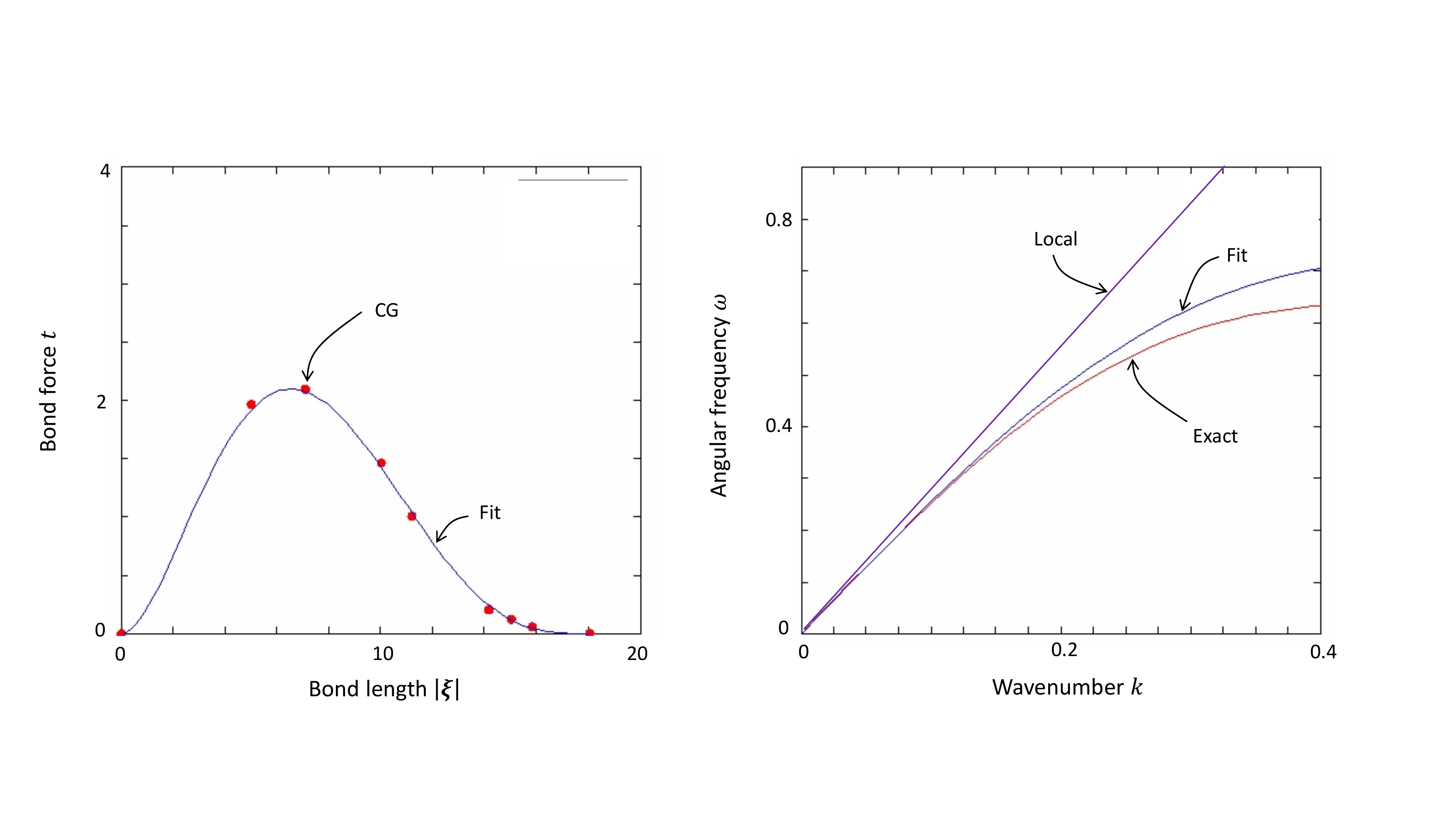}
\caption{Coarse grained and fitted peridynamic model for the square lattice example.
Left: bond force as a function of bond length.
Right: dispersion curves.
}
\label{fig-lrt}
\end{figure}

The small-scale model is deformed in isotropic extension with a strain $\eps$:
\begin{equation}
  \bU_\alpha = \eps\bX_\alpha
\label{eqn-Usqlat}
\end{equation}
where $\eps=0.00019$.
The coarse grained displacements $\bu_i$ and pairwise bond forces $\bff_{ji}$
are evaluated from \eqref{eqn-udef} and \eqref{eqn-ffdef}.
It is convenient to express these forces as being comprised of contributions $\bt_{ji}$ and $\bt_{ij}$
from the material models applied at $i$ and $j$ respectively:
\begin{equation}
  \bff_{ji}=(\bt_{ji}-\bt_{ij})V^2, \qquad \bt_{ji}=-\bt_{ij}=\frac{\bff_{ji}}{2V^2} 
\label{eqn-ft}
\end{equation}
where $V$ is the volume of each coarse grained (CG) node:
\begin{equation}
  V=\tau h^2.
\label{eqn-Vh}
\end{equation}
In this example, $V=1$.
The vector $\bt_{ji}$ is called the {\emph{bond force density}} and has dimensions of force/volume$^2$.
Figure~\ref{fig-lrt} shows the CG bond force densities $|\bt_{ji}|$ 
as a function of CG bond length $|\bxi_{ji}|$ (red dots),
where $\bxi_{ji}=\bx_j-\bx_i$.

In specifying a material model, the bond strain is defined by
\begin{equation}
  s_{ji}=\frac{|\by_j-\by_i|}{|\bx_j-\bx_i|}-1
\label{eqn-sdef}
\end{equation}
where the deformed CG node positions are given by
\begin{equation}
  \by_j=\bx_j+\bu_j
\label{eqn-ydef}
\end{equation}
for any $j$.
Also define the deformed CG bond direction unit vector by
\begin{equation}
  \bM_{ji}=\frac{\by_j-\by_i}{|\by_j-\by_i|}
\label{eqn-Mdef}
\end{equation}
and the normalized bond length by
\begin{equation}
  r_{ji}=\frac{|\bxi_{ji}|}{\delta}.
\label{eqn-rdef}
\end{equation}
For purposes of demonstrating the calibration of a continuum model, suppose a bond-based model is assumed:
\begin{equation}
  \bt_{ji}=\calT_{ji}\bM_{ji}
\label{eqn-tsqlat}
\end{equation}
where $\calT_{ji}$ is a scalar.
The general pattern of the CG bond forces in Figure~\ref{fig-lrt} suggests the following form:
\begin{equation}
  \calT_{ji}=A\sfR(r_{ji})s_{ji}
\label{eqn-trsqlat}
\end{equation}
where
\begin{equation}
  \sfR(r) = r^{\mu_1}(1-r)^{\mu_2}
\label{eqn-Rr}
\end{equation}
and where $A$, $\mu_1$, and $\mu_2$ are constants.
Because the assumed form of the material model \eqref{eqn-trsqlat} is linear in $s_{ji}$ and contains
no dependence on other bonds, it is a bond-based, linear microelastic material model.

To evaluate the parameters, let $i$ be the {\emph{target node}} at the center of the CG grid.
Let $j$ be any node that interacts with $i$.
Taking the logarithm of both sides of each of \eqref{eqn-trsqlat} and rearranging leads to
\begin{equation}
  \log \calT_{ji} = \log A+\mu_1\log r_{ji}+\mu_2\log(1-r_{ji})+\log\eps
\label{eqn-logt}
\end{equation}
where, for uniaxial extension, $s_{ji}=\eps$.
Evaluating $\calT_{ji}$ from the CG data at the three bond lengths $|\bxi_{ji}|=1,2,3$,
\eqref{eqn-logt} forms a linear algebraic system with unknowns $\log A$, $\mu_1$, and $\mu_2$.
This system is easily solved for these quantities.
The parameters $A$, $\mu_1$, and $\mu_2$ are therefore now known.
These values are listed in Table~\ref{table-sqlat}.
  
\begin{table}    
\centering
\begin{tabular}{|l|l|l|l|l|}
\hline
 Parameter             & Value       \\ \hline \hline
 $A$                   & 0.3501      \\ \hline
 $\mu_1$               & 1.902       \\ \hline
 $\mu_2$               & 3.332       \\ \hline
 $\delta$              & 18.03       \\ \hline
\end{tabular}
\caption{Parameters for the peridynamic material model fitted to MD data in the square lattice example.}
\label{table-sqlat}
\end{table}

Figure~\ref{fig-lrt} shows the dispersion curves for the original small-scale model
\eqref{eqn-sqlatt} and the fitted peridynamic model \eqref{eqn-tsqlat}, \eqref{eqn-trsqlat}.
For comparison, the dispersion curve from the local theory (linear elasticity) is also shown.
The peridynamic model provides better agreement with the original model than the local
theory for wavelengths above the CG node spacing.
At smaller wavelengths, the peridynamic model does not include the small-scale interactions that
influence dispersion.
The peridynamic grid has 4\% as many nodes as the original small-scale grid and allows a time step
size 5 times larger.
So, there is a substantially reduced cost in using the coarse grained peridynamic model.

In the continuous form of the peridynamic model, the equation of motion is given by
\begin{equation}
  \rho(\bx)\ddot\bu(\bx,t)=\int_\cHx \big[\bt(\bq,\bx,t)-\bt(\bx,\bq,t)\big]\;\dd\bq +\bb(\bx,t).
\label{eqn-eomt}
\end{equation}
Using \eqref{eqn-trsqlat} and \eqref{eqn-Rr}, the material model in this example problem is then
\begin{equation}
  \bt(\bq,\bx,t)=\bM A\sfR(r)s
\label{eqn-fttsq}
\end{equation}
where
\begin{equation}
  \bM=\frac{\by(\bq,t)-\by(\bx,t)} {|\by(\bq,t)-\by(\bx,t)|}, \quad
  r=\frac{|\bq-\bx|}{\delta}, \quad
  s=\frac{|\by(\bq,t)-\by(\bx,t)|} {|\bq-\by|}-1.
\label{eqn-Msq}
\end{equation}

\section{Application to graphene}\label{sec-graphene}
\begin{figure} 
\centering
\includegraphics[width=1.2\textwidth]{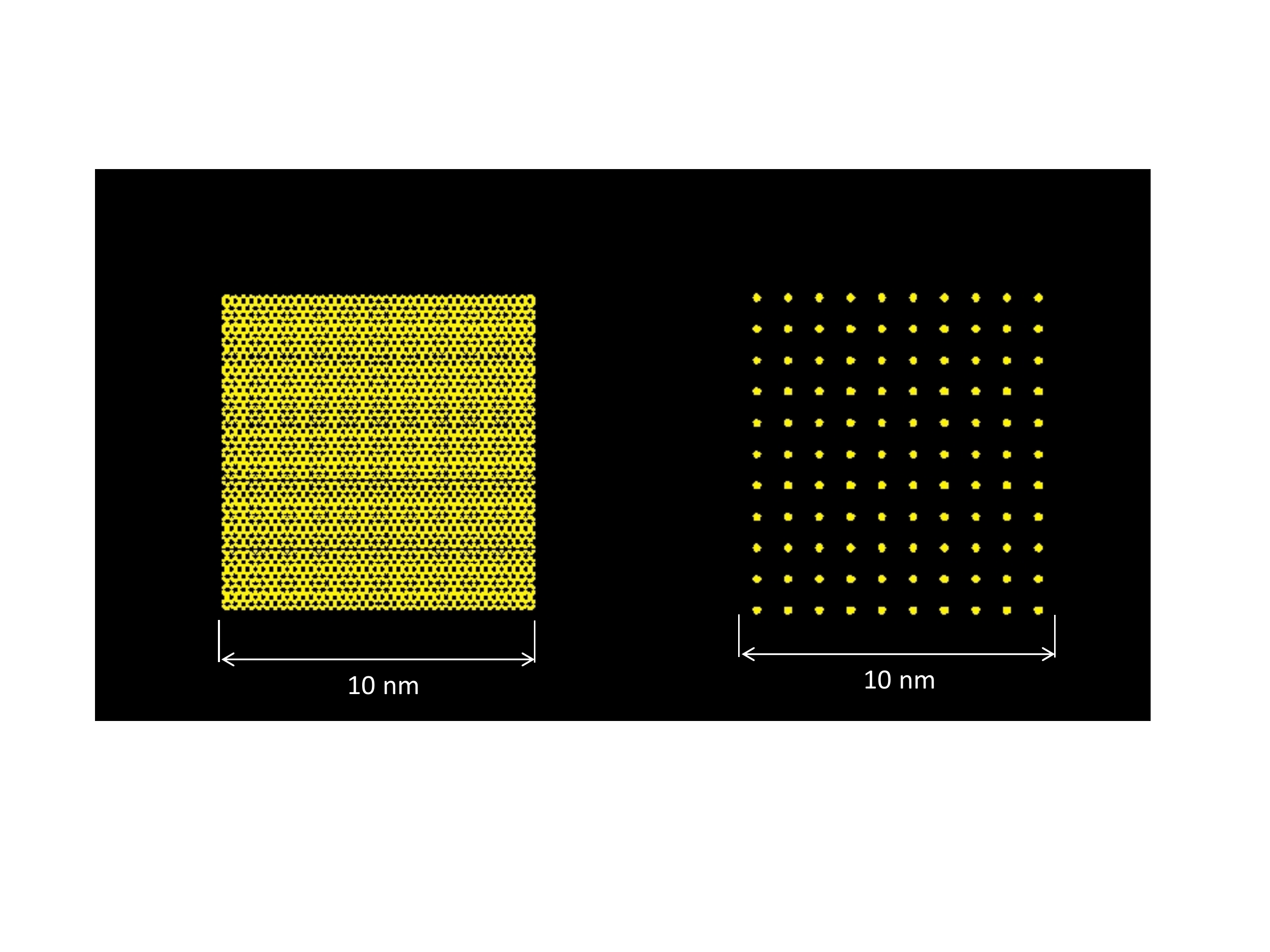}
\caption{Undeformed MD (left) and coarse grained (right) meshes.}
\label{fig-mdmesh}
\end{figure}

To apply the method to graphene, an MD model of a
single-layer graphene sheet was constructed (Figure~\ref{fig-mdmesh}).
The MD mesh is a 10nm square containing 3634 atoms arranged in a hexagonal lattice.
The initial interatomic spacing is 0.146nm.
The atoms interact through a Tersoff potential \cite{tersoff88}.
The temperature is controlled by a thermostat using Langevin dynamics that randomly increases or reduces the
thermal energy of the atoms to keep the mean kinetic energy constant.
To reduce the effect of thermal oscillations on the coarse grained displacements,
the atomic displacements are smoothed over time according the following expression:
\begin{equation}
  \bU_\alpha(0)=0, \qquad \dot\bU_\alpha(t)=\vareps(\tilde\bU_\alpha(t)-\bU_\alpha(t))
\label{eqn-Usmooth}
\end{equation}
where $\tilde\bU_\alpha$ is the unsmoothed displacement of atom $\alpha$ (including thermal oscillations).
The parameter $\vareps$ is a constant taken to be $\vareps=0.005/\Delta t$, where $\Delta t$ is the
MD time step size.
The smoothed displacements $\bU_\alpha$ are used in the coarse grained expressions such as \eqref{eqn-udef}.
The MD grid is initially allowed to reach a constant temperature in an unstressed state before
loading is applied.
After this initial period, constant velocity boundary conditions are applied at the edges of the grid.
When this transition occurs, a velocity gradient is added to the thermal velocities in the grid such
that the atomic velocities are consistent with the boundary conditions.
The thermostat continues to be applied during loading, since otherwise the temperature would change
due to thermoelasticity.

The edges of the MD mesh have prescribed velocity.
The calculation is stopped when the strain exceeds 30\%, at which point the maximum stress
has been reached and the stress is decreasing.
The loading rate is such that this global strain is attained in about 5000 time steps.
To calibrate the peridynamic material model described below, only two loading cases are needed.
These are (1) uniaxial strain, and (2) isotropic extension.

The coarse graining positions $\bx_i$ are generated on a square lattice with spacing $h=0.5$nm.
The weighting functions are the cone-shaped functions given by
\eqref{eqn-Sdef} and \eqref{eqn-Ommd}.
The CG mesh contains 121 nodes. 
Thus, each CG node represents nominally $3634/121\approx30$ atoms.
The CG displacements are computed according to \eqref{eqn-udef}, and the CG bond forces are
computed from \eqref{eqn-ffdef}, using the MD displacements and forces.

\begin{figure} 
\centering
\includegraphics[width=1.2\textwidth]{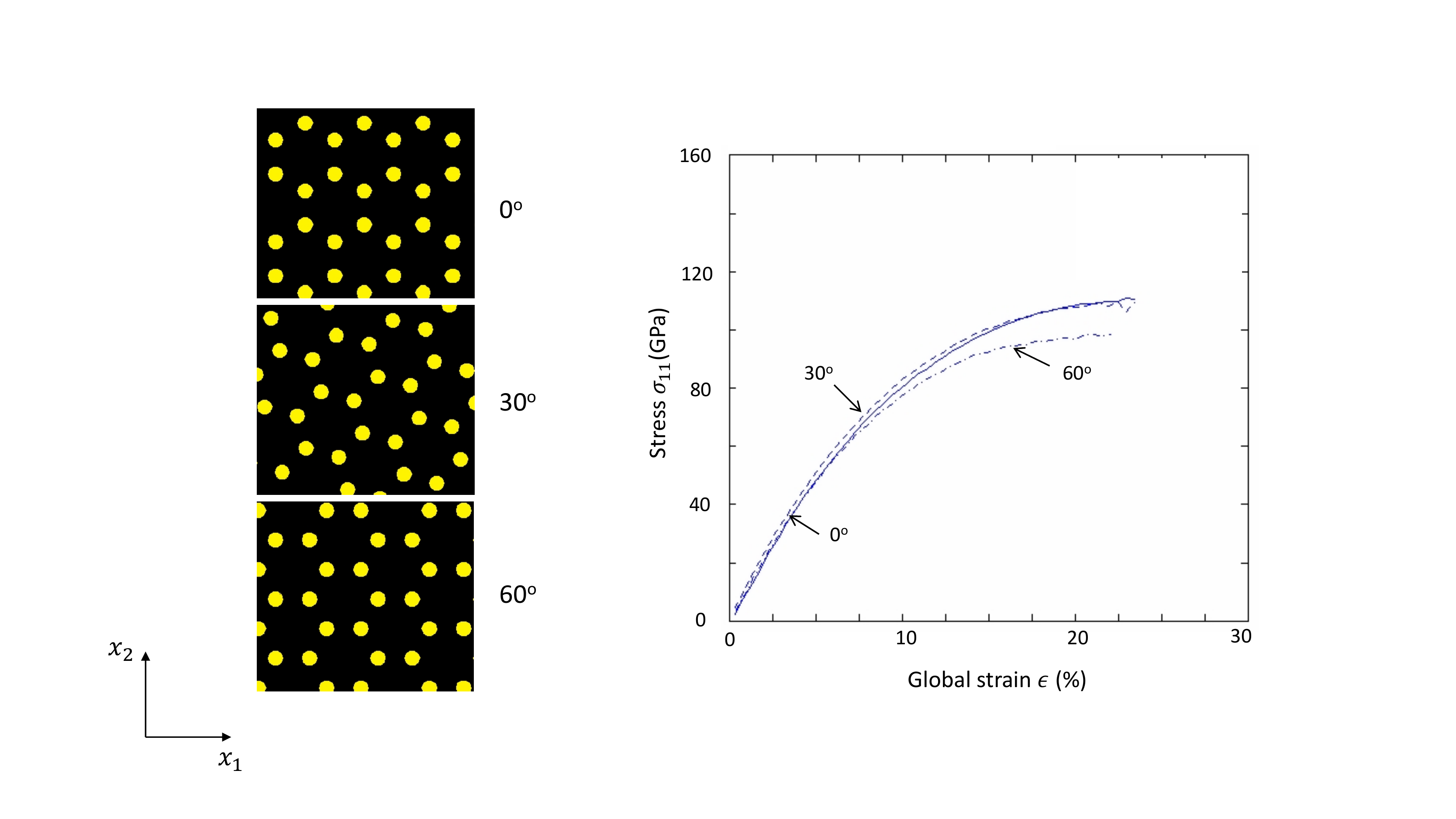}
\caption{Stress-strain curves for graphene under uniaxial strain for three different lattice orientations.}
\label{fig-isotropic}
\end{figure}

The CG bond force data show a softening trend as a function of strain, as shown in
Figure~\ref{fig-isotropic}.
Graphene sheets can be treated as nearly isotropic for purposes of deformation in the plane,
with a significant Poisson effect.
To show this, MD calculations of uniaxial strain at a temperature of 300K were performed with three
different orientations of the hexagonal lattice (Figure~\ref{fig-isotropic}).
The stress-strain curves show that even in the nonlinear regime, the orientation makes only about
a 12\% difference in the stress.

\begin{figure} 
\centering
\includegraphics[width=1.2\textwidth]{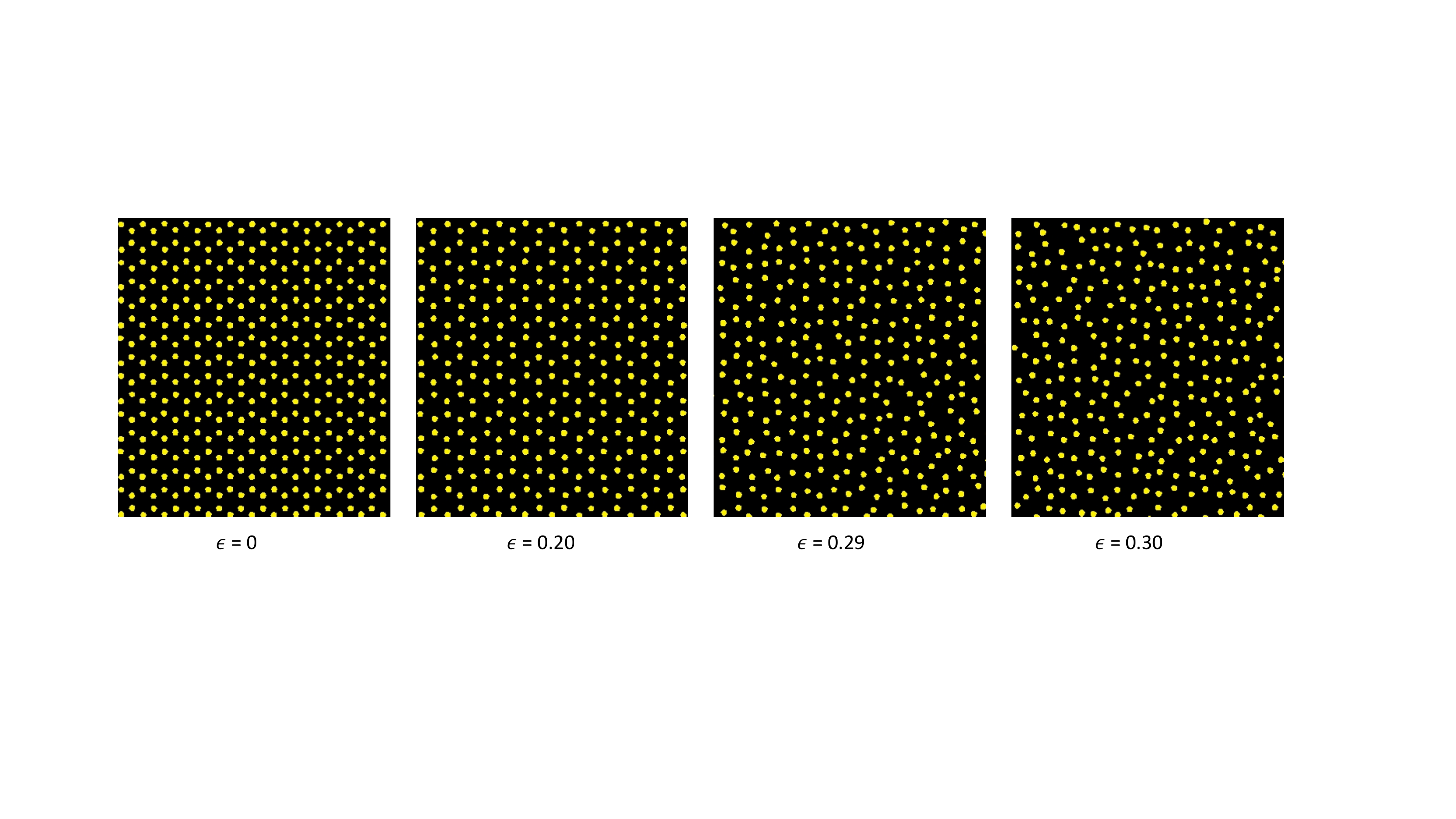}
\caption{MD simulation of uniaxial strain at a temperature of 300K.}
\label{fig-mddeform}
\end{figure}

The process of failure in a typical MD simulation is shown in Figure~\ref{fig-mddeform}.
The graphene sheet at 300K is deformed under (globally) uniaxial strain. 
When the grid is strained beyond the maximum in the stress-strain curve, the perfect hexagonal symmetry
is disrupted due to the onset of material instability, leading rapidly to material failure.

\begin{figure} 
\centering
\includegraphics[width=1.2\textwidth]{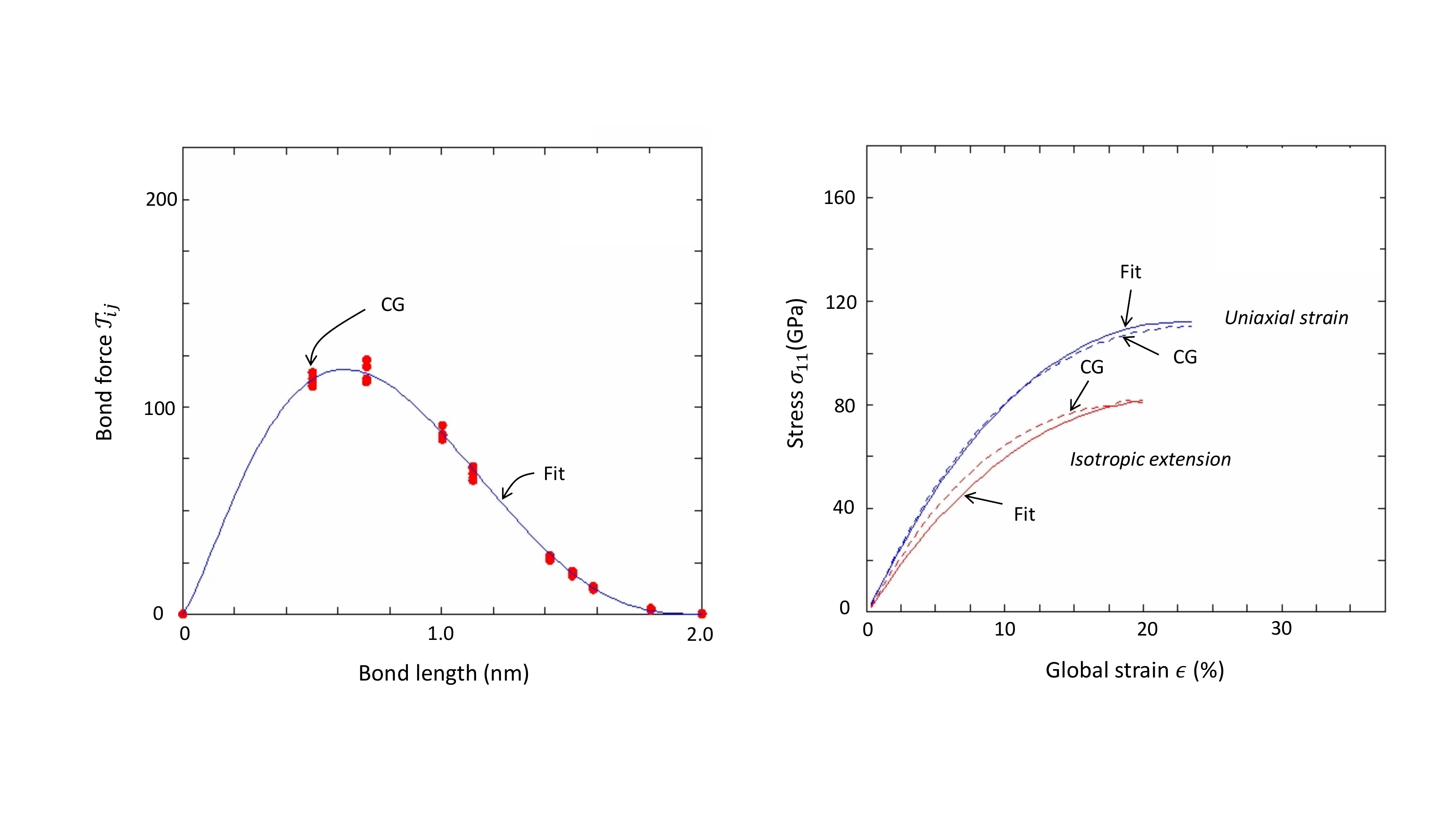}
\caption{Fitted peridynamic material model for coarse grained bond forces in a perfect graphene sheet at 300K.
Left: Dependence of bond force on bond length.
Right: Dependence of bond force on bond strain for a bond with length $h$ in the $x_1$-direction.
}
\label{fig-bondfit}
\end{figure}

To carry out the fitting of a peridynamic model to the CG data, a target CG node $i$
is chosen at the center of the CG mesh.
For node $i$, let $\cH_i$ denote the {\emph{family}} of $i$, defined by
\begin{equation}
  \cH_i=\left\{ j\; \big\vert\; |\bx_j-\bx_i|\le\delta \right\}
\label{eqn-Hidef}
\end{equation}
where $\delta$ is the coarse grained horizon given by \eqref{eqn-deltadef}.
The two MD calculations (for uniaxial strain and isotropic extension), after coarse graining,
provide curves of bond force density $\bt_{ji}$ as a function of the bond strain $s_{ji}$
defined by \eqref{eqn-sdef}.
Also recall the normalized bond length given by \eqref{eqn-rdef}.
Plotting the curves of $|\bt_{ji}|$ as a function of $s_{ji}$ and of $r_{ji}$ for many bonds reveals the general 
shapes shown in Figure~\ref{fig-bondfit}.
The softening response shown in the CG bond forces (dashed lines) suggests the following form:
\begin{equation}
  \calT_{ji}=A\sfR(r_{ji}) \sfS(s_i^+)
     \left[\left(1-\frac{\beta}{2}\right)s_{ji}+\beta \bar s_i\right],
\label{eqn-fasp}
\end{equation}
where the bond length term $\sfR$ has the same form as in the previous example \eqref{eqn-Rr},
and the strain softening term $\sfS$ is given by
\begin{equation}
  \sfS(p) = \left\lbrace \begin{array}{ll}
   n/s_0   & {\mathrm{if}}\;  p\le0,  \\
  \\
  \displaystyle{ 
    \frac{1}{p} \left(1-\left\vert 1-\frac{p}{s_0}\right\vert^n\right) 
               }   & {\mathrm{if}}\; 0<p<2s_0, \\
  \\
    0  & {\mathrm{if}}\;  2s_0<p  \\ 
  \end{array}  \right.
\label{eqn-Ss}
\end{equation}
for any $p$.
The parameters $A$, $\mu_1$, $\mu_2$, $n$, $s_0$, and $\beta$ are constants independent of the bond and of the deformation.
In \eqref{eqn-fasp}, the variables $\bar s_i$ and $s_i^+$ are the mean and maximum strains among all the bonds
in the family of $i$:
\begin{equation}
  \bar s_i = \frac{ \sum_{j\in\cH_i}  s_{ji} }
                  { \sum_{j\in\cH_i} 1 },  \qquad
  s_i^+ = \max_{j\in\cH_i}\big\{ s_{ji}\big\}.
\label{eqn-spm}
\end{equation}
The mean bond strain is similar to a nonlocal dilatation.
In \eqref{eqn-fasp}, the term involving $\beta$ represents the bond strain adjusted by the mean strain.
This term captures the Poisson effect.
The function $\sfS$ is a softening term, which, under tension, drops off to 0 for large strain.
If $\sfS$ were constant, the model would be linearly elastic with variable Poisson ratio.
$\sfS$ depends only on the maximum current bond strain in the family, $s_i^+$.

The next step is to find the parameters in the expressions \eqref{eqn-fasp}--\eqref{eqn-Ss}.
In the following discussion, the stress tensor obtained from the CG bond force data \cite{sill:15} is 
defined by
\begin{equation}
  \bsigma_i = \sum_{j\in\cH_x} \bt_{ji}\otimes\bxi_{ji} V.
\label{eqn-sigmacg}
\end{equation}
The 11 components of the stress tensor in \eqref{eqn-sigmacg} will be denoted by $\sigma_i$ in the present discussion:
\begin{equation}
  \sigma_i(\eps) = \sum_{j\in\cH_x} (t_1)_{ji}(\xi_1)_{ji} V.
\label{eqn-sigma1cg}
\end{equation}
The two coarse grained MD simulations used for calibrating the model parameters have the following strains:
\begin{itemize}
\item Uniaxial strain (UX) with strain $\eps$ in the $x_1$ direction:
\begin{equation}
  s_i^+=\eps, \qquad \bar s_i=\frac{\eps}{2}.
\label{eqn-ux}
\end{equation}
\item Isotropic extension (IE) with strain $\eps$:
\begin{equation}
  s_i^+=\eps, \qquad \bar s_i=\eps.
\label{eqn-ie}
\end{equation}
\end{itemize}
The constant $\beta$ will be determined first.
In the IE and UX cases with global strain $\eps$, the bond strain in a bond with polar angle $\theta$ is given by
\begin{equation}
   s^\IE=\eps, \qquad   s^\UX=\eps\cos^2\theta.
\label{eqn-stheta}
\end{equation}
Then from 
\eqref{eqn-fasp},
\eqref{eqn-sigma1cg}, \eqref{eqn-ux}, \eqref{eqn-ie}, and \eqref{eqn-stheta},
\begin{equation}
   \frac{\sigma_i^\IE}{\sigma_i^\UX}= \frac
       {\sum_{j\in\cH_i} A\sfR(r_{ij})\sfS(\eps) (1+\beta/2)\eps\xi_{ji} \cos^2\theta V}
       {\sum_{j\in\cH_i} A\sfR(r_{ij})\sfS(\eps)[ (1-\beta/2)\cos^2\theta + \beta/2]\eps\xi_{ji} \cos^2\theta V}
\label{eqn-ftheta}
\end{equation}
where $\xi_{ji}=|\bxi_{ji}|$.
Approximating \eqref{eqn-ftheta} by replacing the sums with integrals and noting that
$\sfR$ and $\sfS$ are independent of $\theta$ leads to
\begin{equation}
   \frac{\sigma_i^\IE}{\sigma_i^\UX}= \frac
     {  (1+\beta/2) \int_0^{2\pi}\cos^2\theta\,\dd\theta  }
     {  (1-\beta/2)\int_0^{2\pi}\cos^4\theta\,\dd\theta + (\beta/2) \int_0^{2\pi}\cos^2\theta\,\dd\theta }.
\label{eqn-fthetaint}
\end{equation}
Since $\int\cos^2\theta=\pi$ and $\int\cos^4\theta=3\pi/4$, solving \eqref{eqn-fthetaint} for $\beta$ yields
\begin{equation}
  \beta= \frac{8-6\gamma}{\gamma-4}, \qquad\gamma:= \frac{\sigma_i^\IE}{\sigma_i^\UX}.
\label{eqn-betacal}
\end{equation}
The constants $s_0$ and $n$ are determined next.
For UX, combining \eqref{eqn-fasp}, \eqref{eqn-Ss}, \eqref{eqn-sigma1cg}, and \eqref{eqn-ux} leads to
\begin{equation}
  \sigma_i^\UX(\eps) = \sum_{j\in\cH_x} A\sfR(r_{ji}) \left(1-\left\vert 1-\frac{\eps}{s_0}\right\vert^n\right) (\xi_1)_{ji}V.
\label{eqn-sigmaux}
\end{equation}
The maximum of the function in \eqref{eqn-sigmaux} occurs at $\eps=s_0$, and its value is given by
\begin{equation}
  \sigma_i^\UX(s_0) = \sum_{j\in\cH_x} A\sfR(r_{ji}) (\xi_1)_{ji}V.
\label{eqn-sigmauxmax}
\end{equation}
The values of $s_0$ and $\sigma_i^\UX(s_0)$ are easily read off from the CG data.
Differentiating \eqref{eqn-sigmaux} yields
\begin{equation}
  \frac{\dd\sigma_i^{\UX}}{\dd\eps}(0)=\frac{n}{s_0} \sum_{j\in\cH_x} A\sfR(r_{ji}) (\xi_1)_{ji}V.
\label{eqn-dS}
\end{equation}
The slope of the curve at the origin $\dd\sigma_i^{\UX}/\dd\eps(0)$ is easily obtained from the coarse grained CG
data by numerical differentiation.
Then from \eqref{eqn-sigmauxmax} and \eqref{eqn-dS}, the value of $n$ is found from
\begin{equation}
  n = \frac{s_0}{\sigma_i^{\UX}(s_0)} \frac{\dd\sigma_i^{\UX}}{\dd\eps}(0).
\label{eqn-ns}
\end{equation}
The parameters $s_0$ and $n$ are now known.
The values of $A$, $\mu_1$, and $\mu_2$ are determined from the IE simulation as in Section~\ref{sec-example}
using \eqref{eqn-logt}.
Now all the parameters are known, and the calibration process for the model is complete.
The parameters for the
material model evaluated for the CG node $i$ at the center of the square are given in
Table~\ref{table-matfit}.
A comparison between the fitted peridynamic material model and the coarse grained bond forces
is shown in Figure~\ref{fig-bondfit}.

\begin{figure} 
\centering
\includegraphics[width=1.2\textwidth]{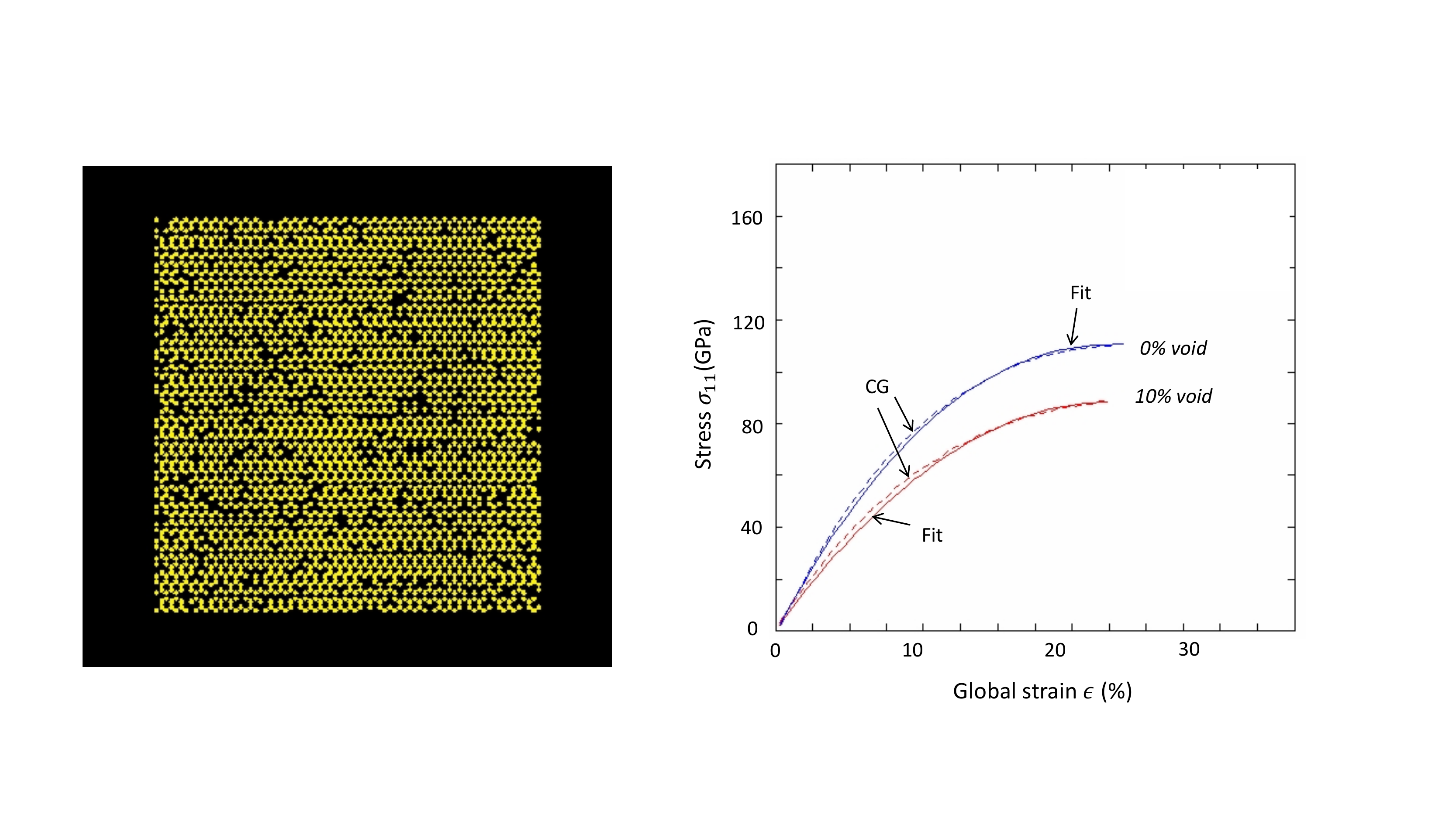}
\caption{Graphene sheet with 10\% void.
Left: Initial MD grid.
Right: CG and fitted peridynamic stress-strain curves in uniaxial strain for 0\% and 10\% void.}
\label{fig-void}
\end{figure}

To illustrate the effect of distributed defects,
the analysis was repeated for a graphene sheet with 10\% of the atoms removed.
The results are shown in Figure~\ref{fig-void}.
As expected, the sample with defects is less stiff and fails at a lower stress.
  
\begin{table}    
\centering
\begin{tabular}{|l|l|l|l|l|}
\hline
 Parameter             & Value         & Units        \\ \hline \hline
 $A$                 & 34.94      & nN/nm$^6$          \\ \hline
 $s_0$                 & 0.2345        &           \\ \hline
 $n$                   & 2.338         &           \\ \hline
 $\mu_1$               & 1.335         &            \\ \hline
 $\mu_2$               & 2.922         &            \\ \hline
 $\beta$               & -1.035        &            \\ \hline
 $\delta$              & 2.121         & nm           \\ \hline
 $G_c$                 & 17.5          & J/m$^2$       \\ \hline
 $s_c$                 & 0.145         &               \\ \hline
 $\tau$                & 0.335         &   nm          \\ \hline
\end{tabular}
\caption{Parameters for the peridynamic material model fitted to MD data at 300K with $h=0.5nm$.}
\label{table-matfit}
\end{table}

The continuous form of the model is then
\begin{equation}
  \bt(\bq,\bx,t)= \bM \calT, \quad
  \calT= A\sfR(r) \sfS(s^+) \left[\left(1-\frac{\beta}{2}\right)s+\beta \bar s\right]
\label{eqn-grcontin}
\end{equation}
where $\bM$ and $r$ are given by \eqref{eqn-Msq} and
\begin{equation}
  s^+(\bx,t)=\max_{\bxi\in\cH} s(\bxi,t), \quad
  \bar s(\bx,t)=\frac{\int_\cH s(\bxi,t)\;\dd\bxi}{\int_\cH \dd\bxi}, \quad
  \bxi=\bq-\bx.
\label{eqn-rpcontin}
\end{equation}

\section{Bond breakage}\label{sec-breakage}
The process of coarse graining described above starts with an MD model that does not
contain initiated cracks, although it can contain distributed defects.
The distinction is that after initiation, the  damage near the crack tip evolves in such
a way that the Griffith criterion applies.
This means that a growing crack consumes a definite amount of energy per unit area of
new crack surface.
This energy is a material property called the {\emph{critical energy release rate}}, 
denoted by $G_c$.
So, the nonlinear material model obtained by coarse graining is designed to simulate 
nucleation of damage, but not the details of what happens in the process zone near
a crack that is already present.

\begin{figure} 
\centering
\includegraphics[width=1.2\textwidth]{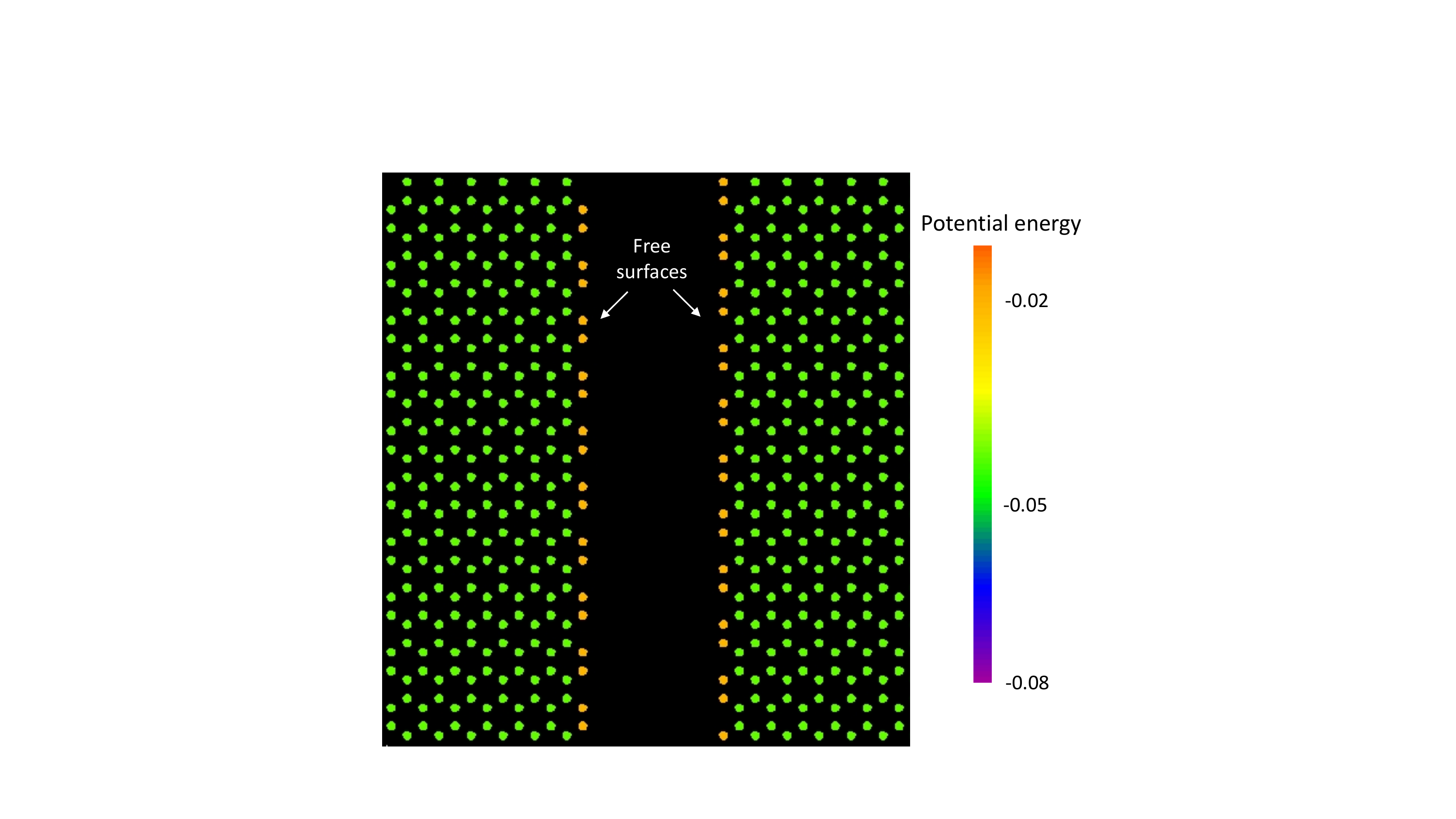}
\caption{Potential energy of atoms on a free edge of a graphene sheet is higher than in the interior.
Colors represent potential energy in the Tersoff interatomic potential.}
\label{fig-esurface}
\end{figure}

To incorporate previously initiated cracks into the continuum model and allow for
rescaling, a value of $G_c$  can be determined easily from the MD model in a separate
simulation.
To do this, assume that all the energy that goes into growing a crack is converted to
surface energy \cite{zhang14}.
The MD interatomic potential is reduced when each atom is surrounded by a certain number 
of neighbors, which is 3 in the case of graphene.
It follows that when some neighbors are removed, as would happen on a crack surface, the
total energy increases.
So, $G_c$ can be determined by performing an MD simulation in which the sample is split into
two halves (Figure~\ref{fig-esurface}).
The total potential energy values before and after the split are $E_0$ and $E_1$ respectively.
The value of $G_c$ is then
\begin{equation}
  G_c=\frac{E_1-E_0}{\tau L},
\label{eqn-GcE}
\end{equation}
where $L$ is the total length of the MD grid along the split and $\tau$ is the thickness (0.335nm
for graphene).
After carrying out the above calculation, the resulting value of $G_c$ is $17.5$J/m$^2$, which is similar
to experimentally measured values \cite{zhang14}.

Bond breakage is added to the coarse grained continuum model \eqref{eqn-grcontin} using the standard
form of irreversible bond breakage:
\begin{equation}
  \calT=A\sfR \sfS\calB(\bxi,t) \left[\left(1-\frac{\beta}{2}\right)s+\beta \bar s\right]
\label{eqn-continbreak}
\end{equation}
where $\calB$ is a binary-valued function that switches from 1 to 0 when the bond $\bxi$ breaks:
\begin{equation}
  \calB(\bxi,t)=\left\lbrace 
    \begin{array}{ll}
      1 & {\mathrm{if}}\;s(\bxi,t')<s_* \;{\mathrm{for\;all}}\;0\le t'\le t, \\
      0 & {\mathrm{otherwise.}}
    \end{array}
  \right.
\label{eqn-mubreak}
\end{equation}
where $s_*$ is the critical strain for bond breakage.
A scalar damage variable $\phi(\bx,t)$ can be defined as the fraction of bonds connected to a point $\bx$ that have broken:
\begin{equation}
  \phi(\bx,t) = 1-\frac{ \int_\cH \calB(\bxi,t)\;\dd\bxi} { \int_\cH \dd\bxi}.
\label{eqn-phidef}
\end{equation}
Once $G_c$ is known from MD, a critical bond strain $s_c$ in the CG material model can be determined by 
requiring that the work per unit area consumed in separating two halves of the CG grid matches
this $G_c$.
Suppose the CG grid is split into two halves $\cR_-$ and $\cR_+$.
Assuming uniaxial strain,
the total work done through the bonds that initially connected the two halves is given by
\begin{eqnarray}
  \tau LG_c&=& \sum_{i\in\cR_-} \sum_{j\in\cR_+}V^2\int_0^{s_c} \calT_{ji}\dd(\xi_{ji}s)   \nonumber\\
  &=&AV^2\sum_{i\in\cR_-} \sum_{j\in\cR_+}\xi_{ji} \sfR(r_{ji}) \int_0^{s_c} \sfS(s)s\;\dd s.
\label{eqn-Es}
\end{eqnarray}
Equation \eqref{eqn-Es} is solved numerically for $s_c$, using the value for $G_c$ that was determined
from MD using \eqref{eqn-GcE}.
Equation \eqref{eqn-Es} is simply the classical expression for the peridynamic energy release rate \cite{madenci:13}
specialized to the present material model.
Since the Griffith fracture criterion only applies to cracks that already exist, rather than new cracks,
the value of $s_c$ obtained from \eqref{eqn-Es} is applied to the bonds connected to $\bx$ only when damage is
already present within the family of $\bx$.
Define the maximum damage within the family of $\bx$ by $\bar\phi(\bx,t)$:
\begin{equation}
  \bar\phi(\bx,t) = \max_{\bq\in\cHx} \phi(\bq,t).
\label{eqn-barphi}
\end{equation}
The critical strain for bond breakage changes from the coarse grained value $s_0$ that reflects crack nucleation
to the Griffith value $s_c$:
\begin{equation}
  s_*=\left\lbrace 
    \begin{array}{ll}
      s_0 & {\mathrm{if}}\;\bar\phi < \phi_{trans} \\
      s_c & {\mathrm{otherwise.}}
    \end{array}
  \right.
\label{eqn-sstar}
\end{equation}
where $\phi_{trans}$ is the transition value of damage, usually set to 0.3.
The use of different values of the critical strain for the nucleation and growth phases is discussed
further in \cite{silling21b} in the context of the microelastic nucleation and growth (MNG) material model.

\section{Changing the horizon}\label{sec-scale}
A peridynamic model obtained from coarse grained data can be rescaled to use any desired horizon $\delta'$.
Let $\delta$ denote the original horizon determined in the coarse graining process, and
let $\kappa=\delta'/\delta$.
It is required that the stress be unchanged by the rescaling:
\begin{equation}
  \int_{\cH'} \bt'(\bxi')\otimes\bxi'\;\dd\bxi' =\int_\cH \bt(\bxi)\otimes\bxi\;\dd\bxi
\label{eqn-resct}
\end{equation}
where $\bt'$ is the rescaled material model, to be determined.
Since the integrals in \eqref{eqn-resct} are area integrals
in 2D, \eqref{eqn-resct} is satisfied for all deformations if $\bt'$ is set to
\begin{equation}
  \bt'(\bxi')=\kappa^{-3} \bt(\bxi'/\kappa)
\label{eqn-rescg}
\end{equation}
for all $\bxi'$.
(In 3D the exponent in \eqref{eqn-rescg} would be $-4$.)

The critical strain derived from the Griffith criterion $s_c$ follows a different scaling relation.
In both 2D and 3D, this relation is given by
\begin{equation}
  s_c'=\kappa^{-1/2} s_c
\label{eqn-ressc}
\end{equation}
which follows from the standard derivation of the critical strain \cite{madenci:13}.
In the application below in Section~\ref{sec-exper}, a value of $\kappa=5$ was used.

The coarse grained material model, before rescaling, embeds length scales from the 
original small scale or MD model, as demonstrated by the dispersion curves in
Figure~\ref{fig-lrt}.
However, these physical length scales are lost when rescaling according to \eqref{eqn-rescg}.
In fact, after rescaling, there may be no compelling reason to use the same bond length dependence
$\sfR$  as was obtained by coarse graining.
This can be replaced by some other convenient form, say $\sfR'$, provided that
\begin{equation}
  \int_0^{\delta'}\sfR'(\xi'/\delta'){\xi'}^2\;\dd\xi'
  =\int_0^{\delta}\sfR(\xi/\delta){\xi}^2\;\dd\xi,
\label{eqn-resR}
\end{equation}
which ensures that the stress is unchanged.

\section{Comparison with experiment}\label{sec-exper}
Lee et al. \cite{lee08} performed experiments in which the elastic response and strength of nearly perfect graphene
sheets were measured.
The sheets were suspended over circular cavities with diameter 1000nm or 1500nm.
The sheets were then deflected by an atomic force microscope (AFM) probe with a nominally hemispherical tip.
The main data reported was the force on the probe as a function of its deflection.

The case with a specimen diameter of 1000nm and an AFM probe tip radius of 27.5nm was simulated with
the coarse grained material model discussed above for a perfect graphene monolayer.
This material model was implemented in the Emu peridynamic code \cite{sias:05}.
The grid spacing in the CG model was scaled up by a factor of 5, resulting in a grid spacing in Emu of 2.5nm
and a horizon of $\delta'=10.61$nm.
The AFM probe tip was modeled as a rigid sphere with constant velocity.

\begin{figure} 
\centering
\includegraphics[width=1.2\textwidth]{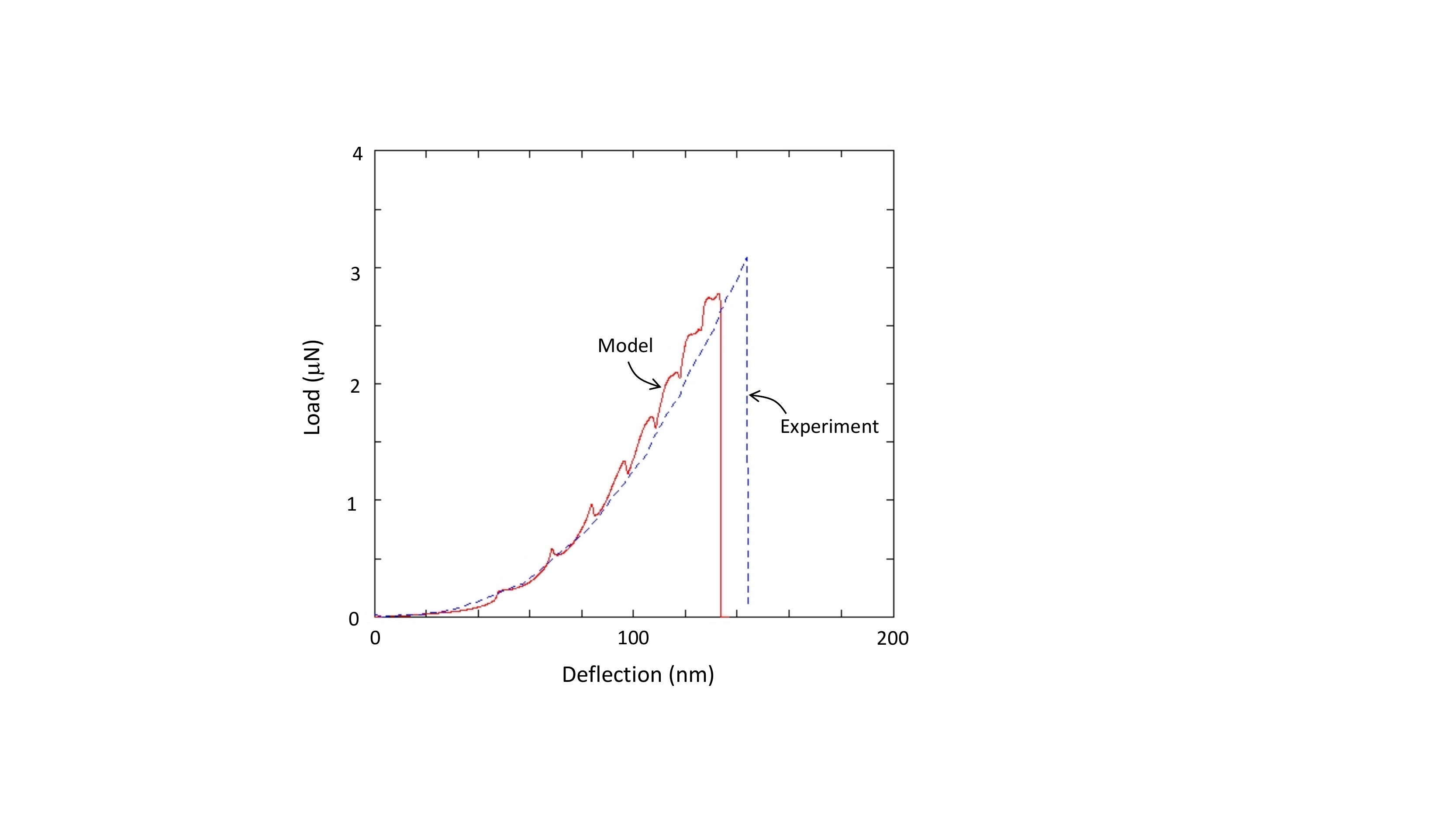}
\caption{Simulated load on an AFM probe deflecting a graphene sheet compared with typical experimental data \cite{lee08}.}
\label{fig-afmload}
\end{figure}

\begin{figure} 
\centering
\includegraphics[width=1.2\textwidth]{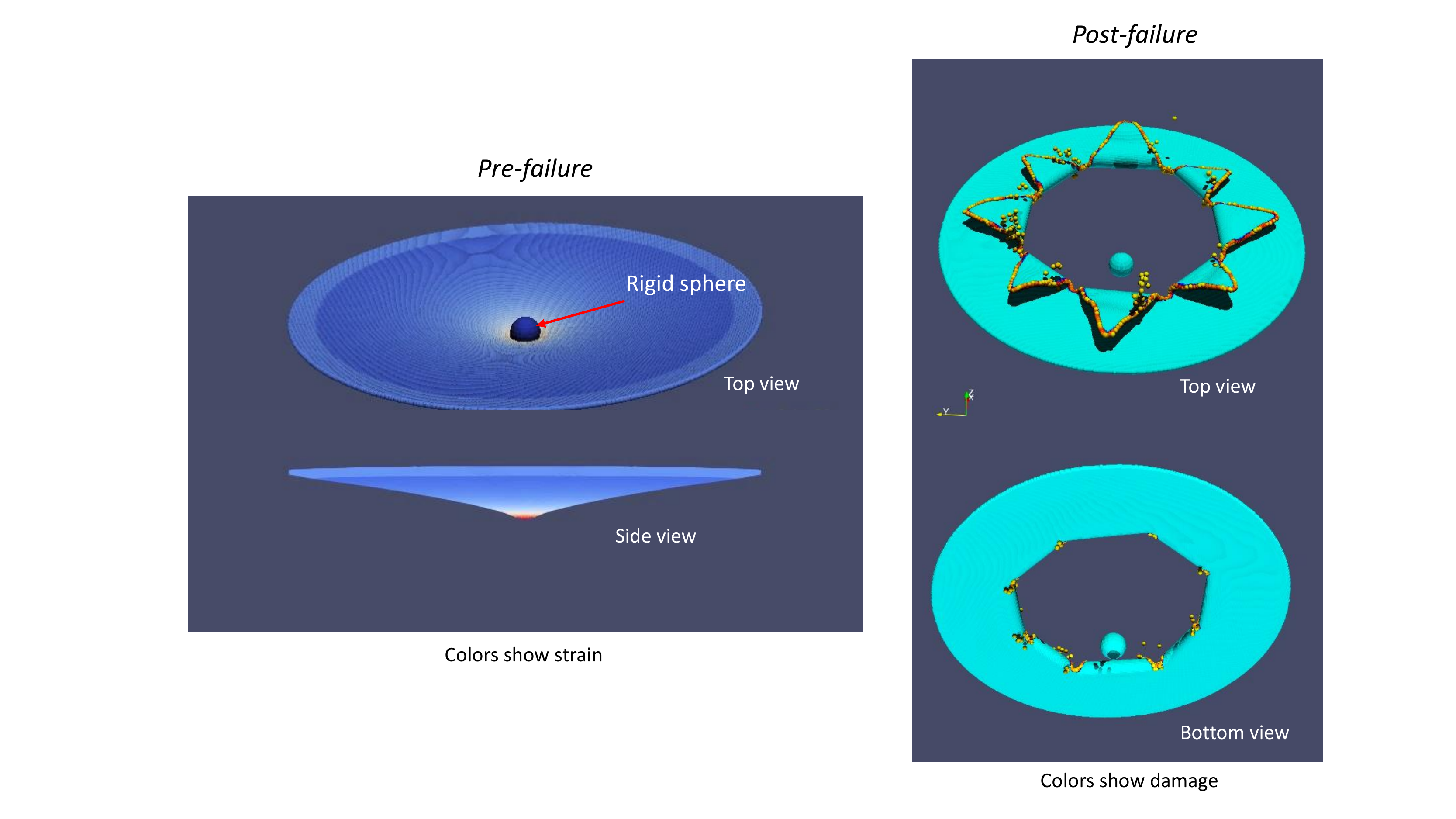}
\caption{Peridynamic simulation of the perforation of a graphene sheet by and AFM probe.}
\label{fig-membrane}
\end{figure}

The load on the AFM predicted by the peridynamic simulation is compared with typical experimental data \cite{lee08} in
Figure~\ref{fig-afmload}.
The experimental data has a statistical variation between tests of about 20\%.
The oscillations in the simulated curve come from vibrations of the membrane in ``trampoline'' mode, since the
simulation is dynamic rather than quasi-static.
The simulation assumed infinite friction, that is, no sliding between the probe and the membrane.
The alternative assumption of zero friction reduces the predicted peak load in the simulation.
It is also uncertain whether the probe is actually hemispherical and smooth, as is assumed in the calculation.
The simulated shape of the membrane and strain distribution just prior to failure are shown in Figure~\ref{fig-membrane}.
After failure, the specimen is predicted to form petals, a feature that is also observed in the experiment.

The Emu calculation had 125,629 nodes and used a time step size of 100fs.
In contrast, a full MD calculation of this problem would require over 28,000,000 atoms and 
have a time step of about 0.5fs.
So, the peridynamic model offers a substantial saving in computer resources compared with full MD.
A peridynamic code with an implicit solver would allow a much larger time step size to be used than in Emu,
which uses explicit differencing in time.

\section{Discussion}\label{sec-disc}
The main result of this paper is a demonstration that the coarse graining method described in Section~\ref{sec-homogpd}
can be used to calibrate an appropriate peridynamic continuum or discretized material model.
The distinguishing features of this method are that it derives nonlocal bond forces directly from MD, and 
that these forces are compatible with the use of smoothed displacements according to a prescribed
weighting function.
A peridynamic material model for graphene obtained from these bond forces
provides good agreement with  nanoscale test data while greatly reducing
the cost of the calculation in comparison with molecular dynamics, especially when used together with
rescaling the horizon.
It was further demonstrated here that the coarse grained model can be combined with standard peridynamic bond
breakage to treat both the nucleation and growth phases of fracture.

As illustrated in Section~\ref{sec-example}, the method can treat long-range forces.
However, graphene sheets do not involve long-range forces, since the Tersoff potential
causes each atom to interact only with its nearest neighbors, of which there are 3.
Long-range forces would arise from the application of surface charge to graphene. 
Long-range forces would also be present in multilayer graphene, since adhesion between the layers
occurs through interactions similar to Van der Waals forces \cite{kiti05}.
So, the capability of the coarse graining method to treat long-range forces
would be needed for these applications.

A possible extension of the method is to apply the calibration process in Section~\ref{sec-graphene} individually
at each CG node, rather than at just one target node $i$.
This would allow the incorporation of defects such as grain boundaries into the calibrated peridynamic model,
in which the material parameters would then become dependent on position.
This extension appears to be practical, because the process of fitting described here is direct, rather than
relying on an optimization technique.

The coarse graining method provides bond forces as the primary quantity that is used for fitting a material
model.
This limits the number of MD simulations that are needed (only uniaxial strain and isotropic extension are used here)
rather than a large suite of training data that might be required in alternative methods.
A different approach \cite{you20,you21,xu21}
is to apply machine learning to fit a peridynamic model to coarse grained displacements.
The machine learning approach avoids the use of coarse grained bond forces but requires many different loading
cases as training data.
Machine learning may offer the potential to learn the form of a peridynamic model from small-scale data
in addition to calibrating the parameters.

\section*{Acknowledgment}
This work was supported by the 
U.S. Army Combat Capabilities Development Command (DEVCOM) Army Research Laboratory 
and by LDRD programs at Sandia National Laboratories.
Sandia National Laboratories is a multimission laboratory managed and operated by 
National Technology and Engineering Solutions of Sandia LLC, a wholly owned subsidiary of 
Honeywell International Inc. for the U.S. Department of Energy’s National 
Nuclear Security Administration under contract DE-NA0003525.
This paper, SAND2021-11007 R, describes objective technical results and analysis. 
Any subjective views or opinions that might be expressed in the paper do not necessarily 
represent the views of the U.S. Department of Energy or the United States Government.

\bibliographystyle{abbrv}
\bibliography{text}

\end{document}